\documentclass[fleqn,usenatbib]{mnras}

\usepackage{newtxtext,newtxmath}

\usepackage[T1]{fontenc}

\DeclareRobustCommand{\VAN}[3]{#2}
\let\VANthebibliography\thebibliography
\def\thebibliography{\DeclareRobustCommand{\VAN}[3]{##3}\VANthebibliography}

\usepackage{graphicx}	
\usepackage{amsmath}	

\title[Time-dependent metal ionization]{Time-dependent metal ionization and the persistence of collisionally excited emission lines in the diffuse ionized gas of star forming galaxies}

\author[L. McCallum et al.]{Lewis McCallum,$^{1}$
Kenneth Wood,$^{1}$
Robert Benjamin,$^{2}$
Dhanesh Krishnarao,$^{3}$
Bert Vandenbroucke$^{4}$
\\
$^{1}$ School of Physics and Astronomy, University of St Andrews, North Haugh, St Andrews, KY16 9SS, UK\\
$^{2}$ Department of Physics, University of Wisconsin-Whitewater, Whitewater, WI 53190, USA\\
$^{3}$ Department of Physics, Colorado College, Colorado Springs, CO 80903, USA\\
$^{4}$  Leiden Observatory, Leiden University, PO Box 9513, 2300 RA Leiden, the Netherlands
}

\date{Accepted XXX. Received YYY; in original form ZZZ}

\pubyear{2024}

\begin{document}
\label{firstpage}
\pagerange{\pageref{firstpage}--\pageref{lastpage}}
\maketitle

\begin{abstract}
We extend our time-dependent hydrogen ionization simulations of diffuse ionized gas to include metals important for collisional cooling and diagnostic emission lines. The combination of heating from supernovae and time-dependent collisional and photoionization from midplane OB stars produces emission line intensities (and emission line ratios) that follow the trends observed in the Milky Way and other edge-on galaxies. The long recombination times in low density gas result in persistent large volumes of ions with high ionization potentials, such as \ion{O}{III}  and \ion{Ne}{III}. In particular, the vertically extended layers of \ion{Ne}{III}  in our time-dependent simulations result in [Ne III] 15$\mu$m/[Ne II] 12$\mu$m emission line ratios in agreement with observations of the edge-on galaxy NGC~891. Simulations adopting ionization equilibrium do not allow for the persistence of ions with high ionization states and therefore cannot reproduce the observed emission lines from low density gas at high altitudes.

\end{abstract}

\begin{keywords}
methods: numerical -- HII regions -- ISM: structure -- ISM: kinematics and dynamics -- galaxies: ISM -- galaxies: star formation
\end{keywords}



\section{Introduction}

The diffuse ionized gas (DIG) in galaxies, a pervasive component of the interstellar medium (ISM), has been an active area of research interest following its initial detection in the Milky Way by \citet{hoyleellis} and subsequent confirmation through $\rm{H}\alpha$ observations \citep{reynolds73}. Extensive surveys have mapped the Milky Way and other galaxies in $\rm{H}\alpha$ and collisionally excited emission lines, unveiling the characteristics of the DIG including its scaleheight, temperature, and ionization state (\citet{haffner03}; \citet{gaensler08}; \citet{hill08}; \citet{krishnarao17}). A critical aspect for understanding the DIG involves identifying the mechanisms supporting and ionizing this gas at high altitudes, with many studies indicating the central role of supernovae and ionizing photons in driving galactic turbulence and outflows (\citet{dealv05}; \citet{joung09}; \citet{hill12}).

Massive O and B stars, with their substantial ionizing luminosity, are believed to be significant contributors to the ionization of the DIG (\citet{reynolds90}; \citet{miller93}; \citet{dove94}). A possible additional source of ionization, particularly relevant at higher altitudes, is photoionization from hot low mass evolved stars (HOLMES). These stars, while less luminous than OB stars, are more numerous, exist at higher scale heights, and may contribute significantly to the DIG's ionization state (\citet{Byler2019}; \citet{rand11}; \citet{floresfjardo}; \citet{vandenbroucke19}). However, recent theoretical investigations adopting ionization equilibrium reveal that the support and structure of the DIG cannot be solely attributed to ionizing photons. For instance, \citet{vandenbroucke19} showed that photoionization alone is insufficient to dynamically maintain the DIG densities observed. Similarly, in the simulations presented by \citet{kadofong} the DIG layer displayed large fluctuations in density and scaleheight, and in general was simulated to have densities and scaleheights which are lower than observed in the Milky Way.

In our preceding work, \citet{mccallum24}, we made progress towards understanding the ionization state of low density gas in the ISM. One of our key contributions was the development and application of a time-dependent ionization calculation. This approach marked a departure from our previous simulations that adopted ionization equilibrium, revealing the dynamic nature of ionization processes in the low density DIG where recombination times can be millions of years and often longer than the lifetimes of OB stars. We demonstrated that the low-density DIG is not in ionization equilibrium, but subject to fluctuations driven by the lifetimes of ionising sources and collisional ionization in gas heated by supernovae. These insights help to explain the ionization and spatial extent of DIG in the Milky Way and other galaxies.

The success of our new simulations in producing the inferred density and scaleheight of DIG in the Milky Way provides a foundation for the work of this paper. By extending our time-dependent ionization calculations of hydrogen to a wider range of astrophysical metals, we will refine our understanding of the ionization and dynamics in the DIG. 

A persistent challenge in simulations of the DIG has been the production of the observed trends in emission line ratios such as [N II]/H$\alpha$, [S II]/H$\alpha$, [O III]/H$\alpha$, [Ne III]/[Ne II]. General trends are that the emission line ratios increase towards higher altitudes (lower density gas) in the Milky Way and other galaxies (\citet{haffner99}; \citet{rand08}; \citet{Jones17}; \citet{levy19}). Despite numerous efforts, radiation-hydrodynamic (rad-hydro) simulations have yet to self-consistently capture the observed trends. The most advanced rad-hydro simulations of the vertically extended ISM to date include those of the SILCC project (\citet{walch15}; \citet{rathjen21}; \citet{rathjen23}) and TIGRESS-NCR \citep{kim23}. While these simulations capture great detail on the dynamics of the ISM, star formation and feedback mechanisms, the ionization states of most metals are not tracked explicitly due to the added computational expense in otherwise highly detailed simulations. Because of this, calculation of emission line intensities can only be done through post-processing techniques which neglect the possibility that much of the high altitude gas may be out out of ionization equilibrium. Emission line intensities and their ratios are sensitive indicators of the ionization and thermal state of the gas, varying with altitude due to changes in density, temperature, and ionizing radiation fields. However, the assumption of ionization equilibrium may oversimplify these dynamic environments, where the availability of ionizing photons can vary faster than the gas is able to respond.

As demonstrated by \citet{mccallum24}, regions where the recombination timescales are longer than the time-variability (or lifetimes) of the ionizing sources is where the equilibrium assumption becomes particularly problematic. The failure to self-consistently replicate the observed emission line trends suggests that we may be missing key aspects of the ionization processes in these regions. This work aims to address this gap by incorporating time-dependent metal ionization calculations, providing a more accurate representation of the ionization states. By doing so, we find insights into the mechanisms governing the ionization state of gas at various altitudes.

\section{Methods}

\subsection{Underlying Radiation-Hydrodynamics Simulations}

\begin{figure}
    \centering
    \includegraphics[width=1.0\columnwidth]{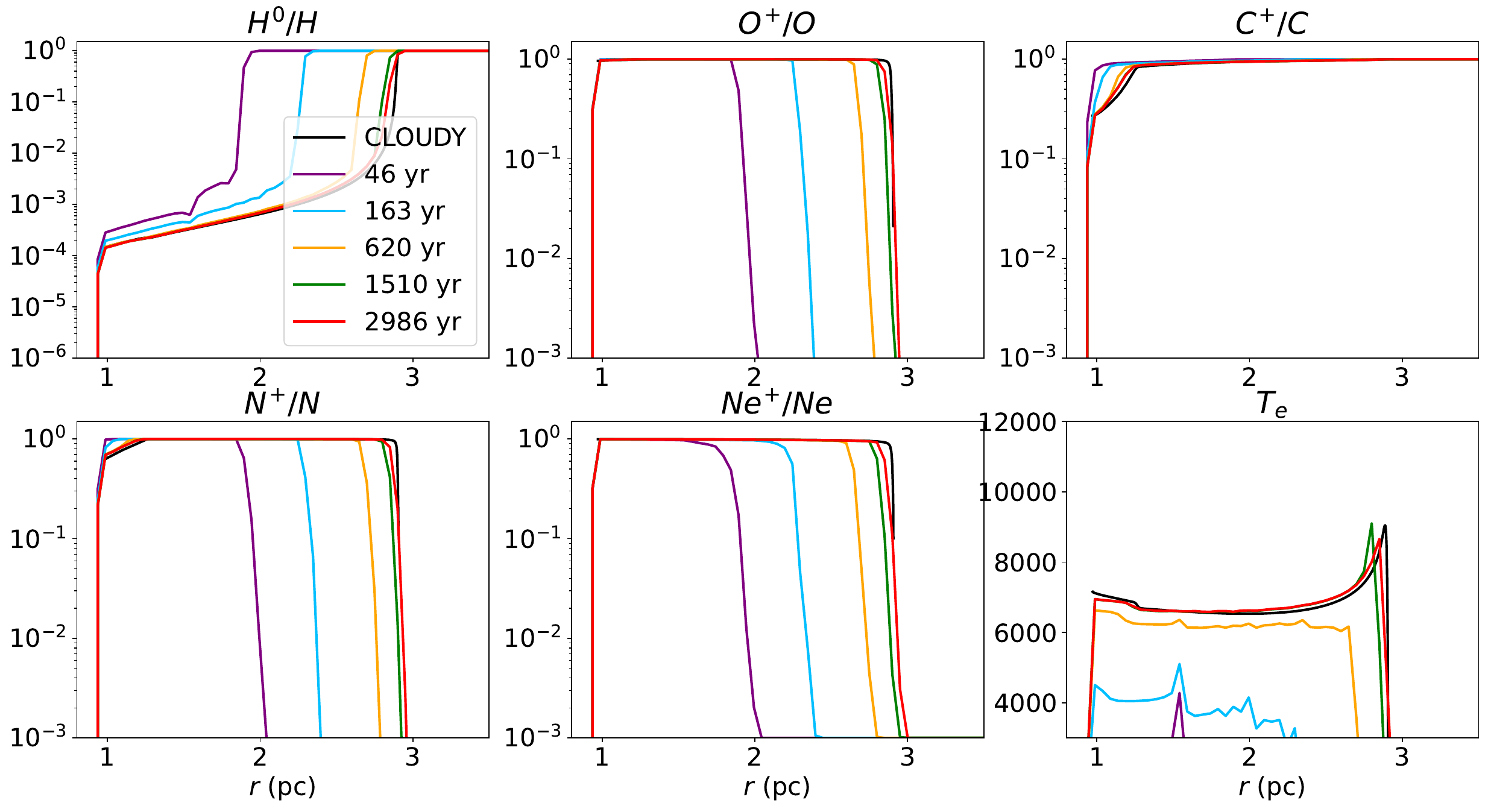}
    \caption{Time dependent evolution of the HII20 Lexington benchmark, with the full cooling calculation. Black line shows equilibrium result from \texttt{CLOUDY}. Coloured lines show a sample of ionization states and temperature at various times. Simulation has stopped evolving by the final coloured line.}
    \label{lex20}
\end{figure}

\begin{figure}
    \centering
    \includegraphics[width=1.0\columnwidth]{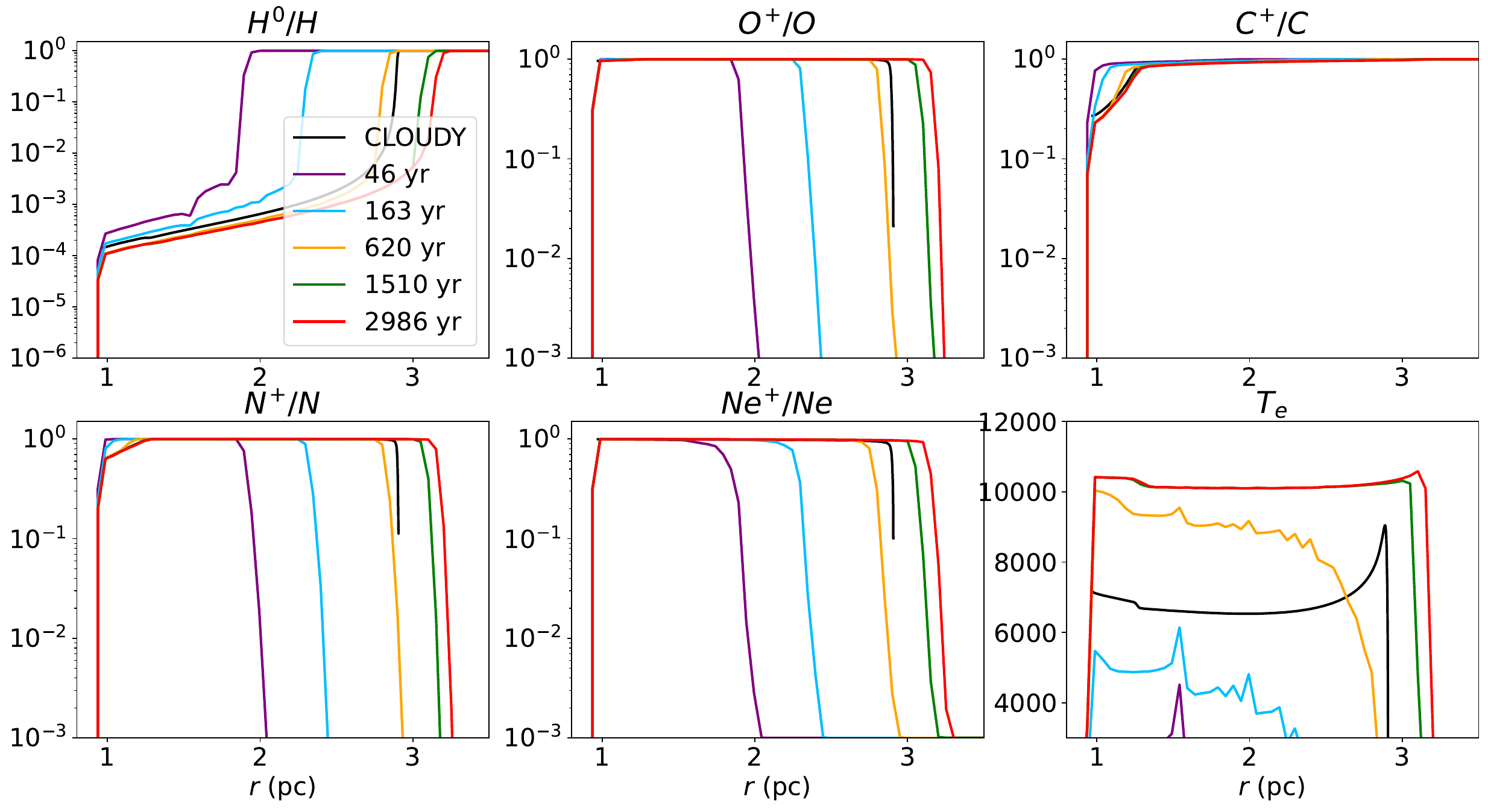}
    \caption{Time dependent evolution of the HII20 Lexington benchmark, with cooling from tabulated cooling curves. Black line shows equilibrium result from \texttt{CLOUDY}. Coloured lines show a sample of ionization states and temperature at various times. Simulation has stopped evolving by the final coloured line.}
    \label{lex20table}
\end{figure}

Our first study was based on hydrogen-only simulations conducted using the Monte-Carlo radiation-hydrodynamics code CMacIonize (\citet{cmacionize}; \citet{cmi2}). The simulations employed a static Cartesian grid measuring 1 × 1 × 6 kpc with 128 × 128 × 768 cells (for a resolution of 7.8~pc), periodic boundary conditions in X-Y axes, allowing outflow in the Z axis.
One new aspect of our simulations was the adoption of a non-isothermal equation of state with a polytropic index of 5/3. This approach integrated heating contributions from supernovae and photoionization, with cooling tables from \citet{derijke13}. Photoionization feedback from massive stars was incorporated through Monte Carlo radiation transfer simulations conducted every 0.2~Myr, using a WMBASIC model stellar spectra \citep{pauldrach01} for OB stars ($T_{\rm{eff}} = 40~kK$), and a Pegase3 stellar population spectrum \citep{fioc19} for HOLMES sources (with an age of 10~Gyr). We do not include the opacity from dust grains in our radiative transfer step, as we expect much of the dust effects to occur in the dense, unresolved molecular clouds, whereby losses of ionizing photons due to dust are contained within our adopted ionizing photon molecular cloud escape fraction.

We implement time-dependent ionization calculations, which is necessary to accurately determine the ionization state of hydrogen in lower density environments. This involved updating the ionization state at each radiation timestep, with a limiter method ensuring a realistic representation of ionization states over time.
Supernova feedback, a vital component in our simulations, was implemented following the methodology of \citet{gatto15}, featuring two modes: direct thermal energy injection and momentum injection. The mode of injection was determined based on the spatial resolution relative to the Sedov-Taylor blast radius.
Star formation within the simulation was categorized into peak driving and random driving. Peak driving represented the formation of extremely massive stars in dense molecular clouds, while random driving is for stars greater than $8M_{\odot}$ that have moved away from denser regions to inject energy into the lower-density ISM. The star formation rate was initially set as an approximate mean Miky Way value, and adjusted to correlate with the evolving surface density of the disc via the Kennicutt-Schmidt relation.
Additionally, the ionizing luminosity from low mass hot evolved stars (HOLMES) and white dwarfs was considered. These sources, although lower in ionizing luminosity, are significant in number and were included in the simulations with varying ionizing luminosities.
The initial conditions were set up to reflect the Dickey-Lockman and Reynolds layer structures of neutral and ionized hydrogen in the Milky Way. An external gravitational potential simulates the gravitational effects from stars and gas within the galaxy, following the approach of \citet{walch15} and \citet{vandenbroucke19}.

The suite of radiation-hydrodynamics simulations as described in \citet{mccallum24} will be used as initial conditions for our new simulations calculating a more precise time-dependent temperature structure and ionization structure of various astrophysical metals, and the associated recombination and collisionally-excited emission line maps. 

Grids are initialised as entirely neutral for all metals at the start of our simulations, with the density, velocity, temperature and hydrogen ionization structure being used from a single snapshot from the hydrogen only simulations. An ionization state settling period of around 10~Myr was found to be sufficient for the simulation to 'forget' its neutral initial conditions. This 10~Myr period was deemed sufficient by running a comparison simulation whereby the grids were initialised in a higher state of ionization, and run to a point where the ionization states were qualitatively similar to those initialised as neutral.

\subsection{Time Dependent Ionization Calculations}

We apply the time-dependent calculation of ionization states throughout each radiation step, where one radiation step is a series of Monte-Carlo radiative transfer simulations. Photon packets are tracked through the grid from source locations until they are either absorbed or escape the grid. After one radiative transfer step, the ionization states of hydrogen and helium are updated in accordance with the timestep, collisional and photoionization rates. With the ionization states of hydrogen and helium updated, the next radiative transfer loop can begin. This iterative process is necessary as the effective opacity in each grid cell is dependent upon the ionization state of hydrogen and helium. While the work of \citet{mccallum24} implemented a limiter method for computational expediency, in this work we solve the following equations using a fourth order Runge-Kutta ODE solver. The rate of change of hydrogen ionization is calculated via equation~\ref{hydrogenion}, and both states of helium are updated by equations~\ref{heliumion1} and \ref{heliumion2}.

\begin{equation}
\begin{aligned}
     \frac{dx_{H^{0}}}{dt} = -\gamma(H^{0}) n_{e} x_{H^{0}} - J_{H} x_{H^{0}} + \alpha(H^{0}) (1 - x_{H^{0}}) n_{e} \\ - n_{e}P(H_{OTS})\alpha^{eff}_{2^{1}P} A_{He} x_{He^{+}}
    \label{hydrogenion}
\end{aligned}
\end{equation}

\begin{equation}
    \frac{dx_{He^{0}}}{dt} = -\gamma(He^{0})n_{e}x_{He^{0}} - J_{He^{0}}x_{He^{0}} + \alpha(He^{0})n_{e}x_{He^{+}}
    \label{heliumion1}
\end{equation}

\begin{equation}
\begin{aligned}
     \frac{dx_{He^{+}}}{dt} = \gamma(He^{0})n_{e}x_{He^{0}} + J_{He^{0}}x_{He^{0}} - \alpha(He^{0})n_{e}x_{He^{+}} \\ - \gamma(He^{+})n_{e}x_{He^{+}} + \alpha(He^{+})n_{e}x_{He^{++}}
    \label{heliumion2}
\end{aligned}
\end{equation}

where $A_{He}$ is the abundance of helium, $\gamma$ is a collisional ionization rate, $\alpha$ is a recombination rate, $J$ is the photoionization mean intensity which intersects the subscripted species, $x$ is the ionic fraction of the subscripted species, $n_{e}$ is the electron density, $P(H_{OTS})$ is the probability a helium Lyman-$\alpha$ photon is absorbed on the spot by a neutral hydrogen, and $\alpha^{eff}_{2^{1}P}$ is the rate of the transition leading to this photon.
 
After 10 iterations, the ionization states of hydrogen and helium are seen to be converged, and the electron number density calculated is used to carry out the time-dependent ionization calculation for the other metals. This calculation is only carried out after the iterative steps. Due to the low abundances of metals, we assume they do not contribute to the electron number density or the opacity for ionizing photons. 

For any ionization state of any metal, the general form of the equation for the rate of change of ionization state is as shown in equation~\ref{metaleqn}. For the ground level, the coefficients for transitions to and from $n-1$ are zero.

\begin{equation}
\begin{aligned}
\frac{dx_{M_{n}}}{dt} = -\gamma(M_{n})x_{M_{n}}n_{e} - J_{M_{n}}x_{M_{n}} + \alpha(M_{n})x_{M_{n+1}}n_{e} \\ + \gamma(M_{n-1})x_{M_{n-1}}n_{e} + J_{M_{n-1}}x_{M_{n-1}} - \alpha(M_{n-1})x_{M_{n}}n_{e} \\ - \beta(M_{n})x_{M_{n}} + \epsilon(M_{n})x_{M_{n}} + \beta(M_{n-1})x_{M_{n-1}} \\ - \epsilon(M_{n-1})x_{M_{n-1}}
    \label{metaleqn}
\end{aligned}
\end{equation}

The terms are the same as defined for equations~\ref{hydrogenion}, \ref{heliumion1} and \ref{heliumion2}, with the addition of $\epsilon$ and $\beta$, the total charge transfer recombination and ionization rates respectively.

Beyond hydrogen and helium, the metals tracked by our code in these simulations are as follows; C, N, O, Ne, S.

The highest ionization state of each metal is not tracked explicitly, as we assume the remaining abundance of each metal is in that state.

Not every possible ion is tracked, with the highest ionization state calculated representing gas which is either in the highest calculated ionization state or higher. For example we calculate oxygen ionization up to \ion{O}{V} , with gas marked in the \ion{O}{V}  state being in one of any state between \ion{O}{V}  and \ion{O}{IX}.

For carbon and sulphur the lowest ionization state is assumed to be the singly ionized state (\ion{C}{II}  and \ion{S}{II} ), due to the low ionization energies of \ion{C}{I}  and \ion{S}{I} .

Equation~\ref{metaleqn} is solved at each timestep using a fourth order Runge-Kutta-Fehlberg solver.

\begin{figure}
    \centering
    \includegraphics[width=1.0\columnwidth]{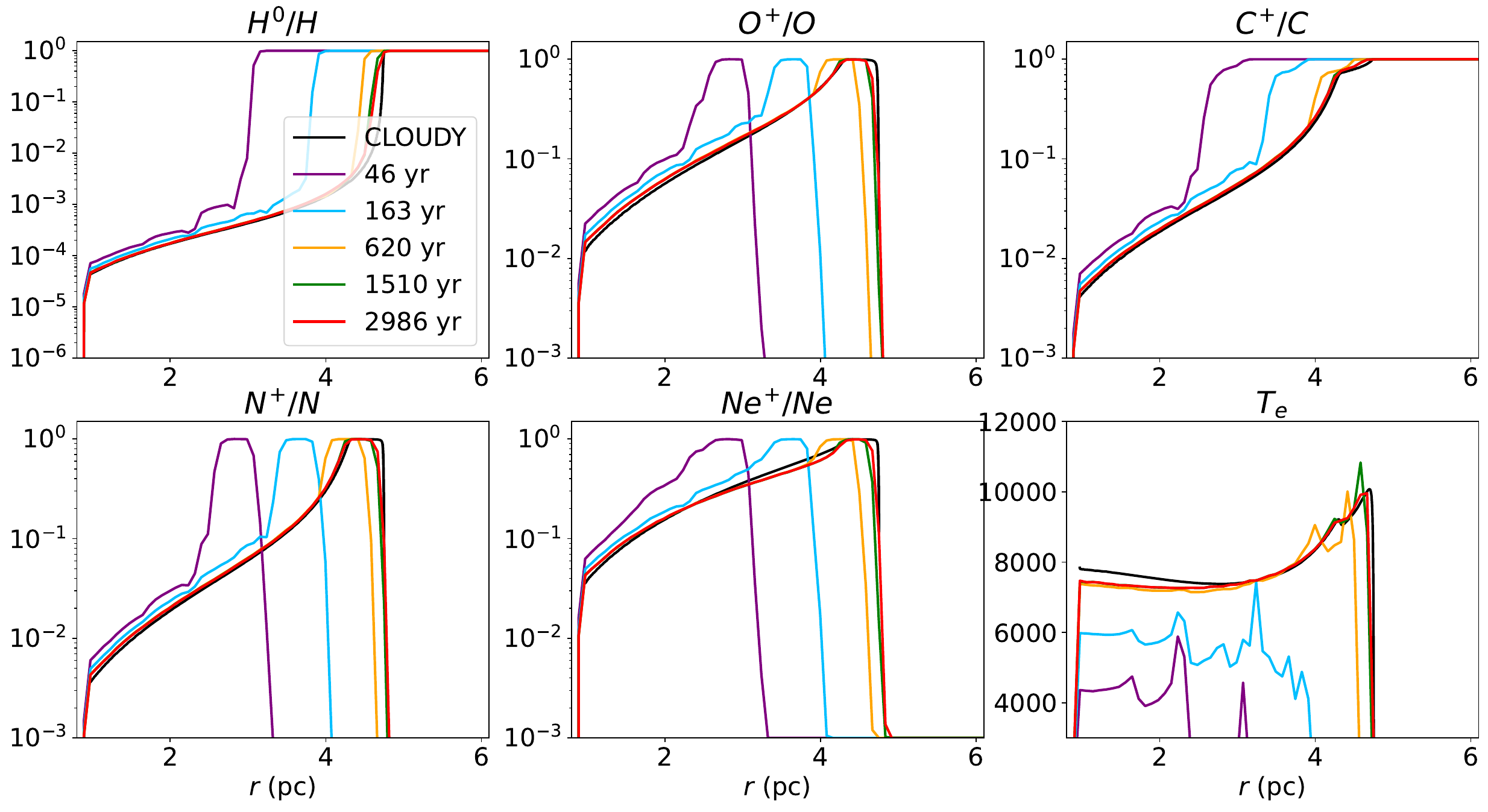}
    \caption{Time dependent evolution of the HII40 Lexington benchmark, with the full cooling calculation. Black line shows equilibrium result from \texttt{CLOUDY}. Coloured lines show a sample of ionization states and temperature at various times. Simulation has stopped evolving by the final coloured line.}
    \label{lex40} 
\end{figure}

\begin{figure}
    \centering
    \includegraphics[width=1.0\columnwidth]{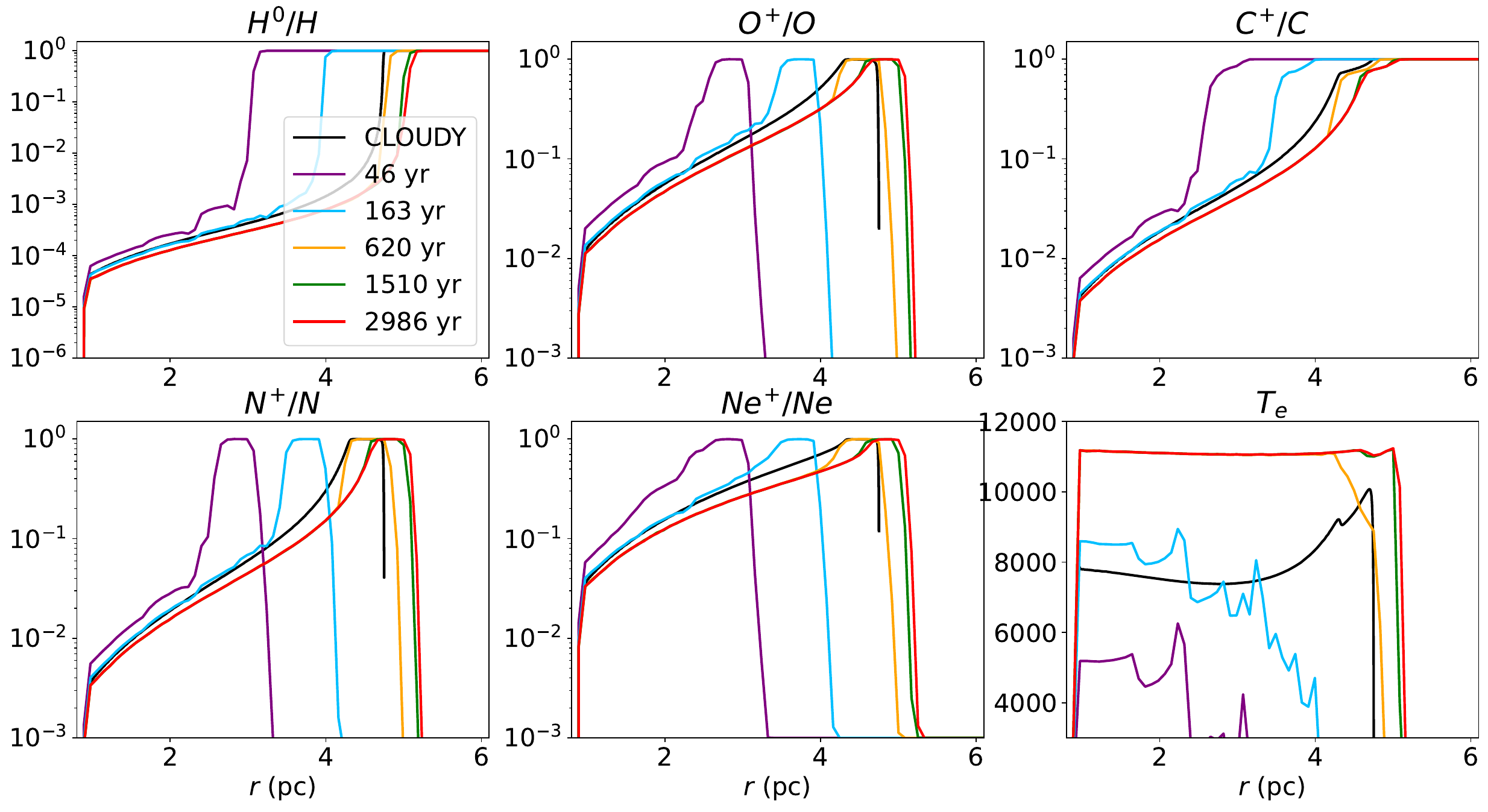}
    \caption{Time dependent evolution of the HII40 Lexington benchmark, with cooling from tabulated cooling curves. Black line shows equilibrium result from \texttt{CLOUDY}. Coloured lines show a sample of ionization states and temperature at various times. Simulation has stopped evolving by the final coloured line.}
    \label{lex40table}
\end{figure}

\begin{figure*}
    \centering
    \includegraphics[width=0.7\textwidth]{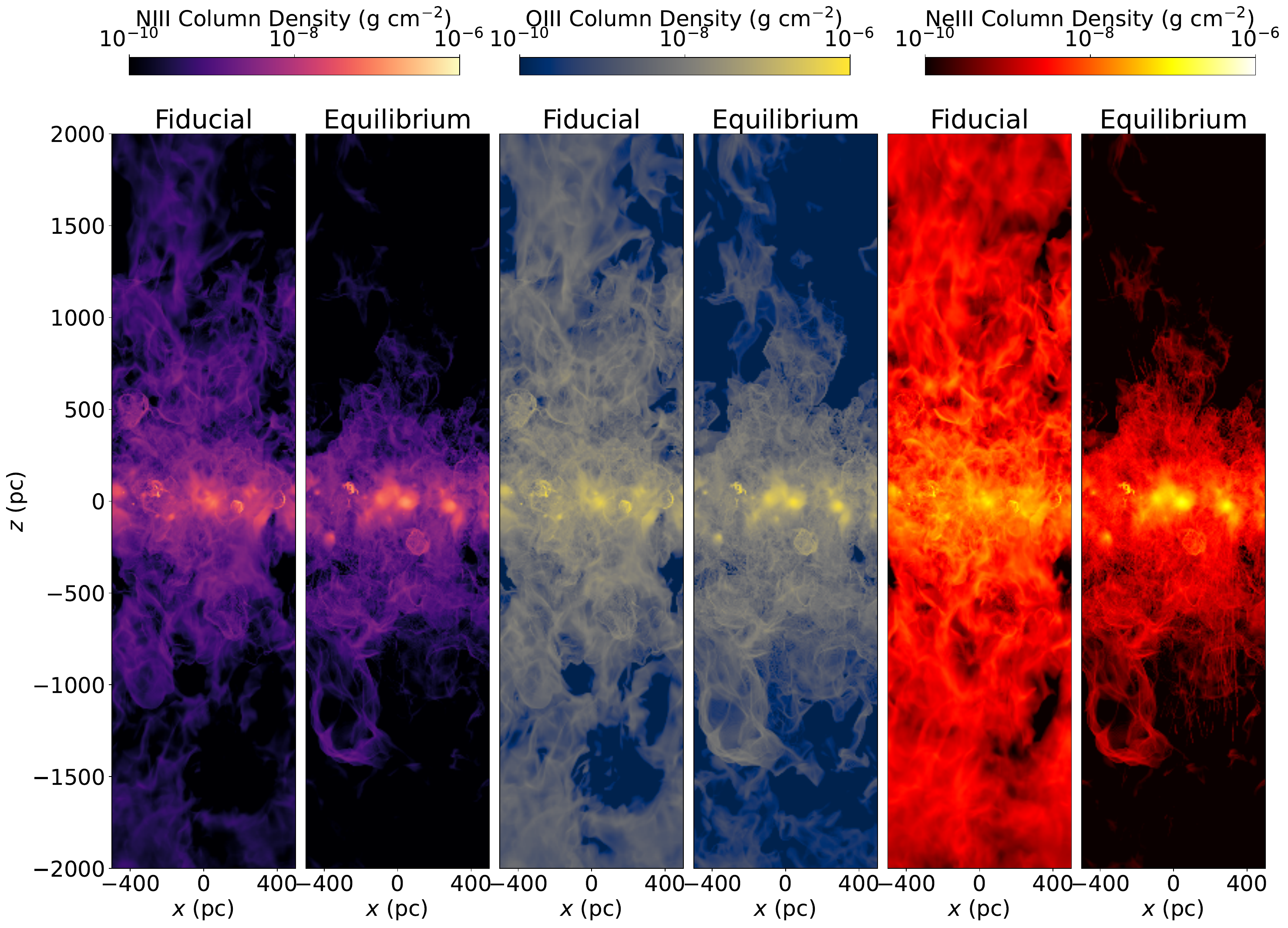}
    \includegraphics[width=0.7\textwidth]{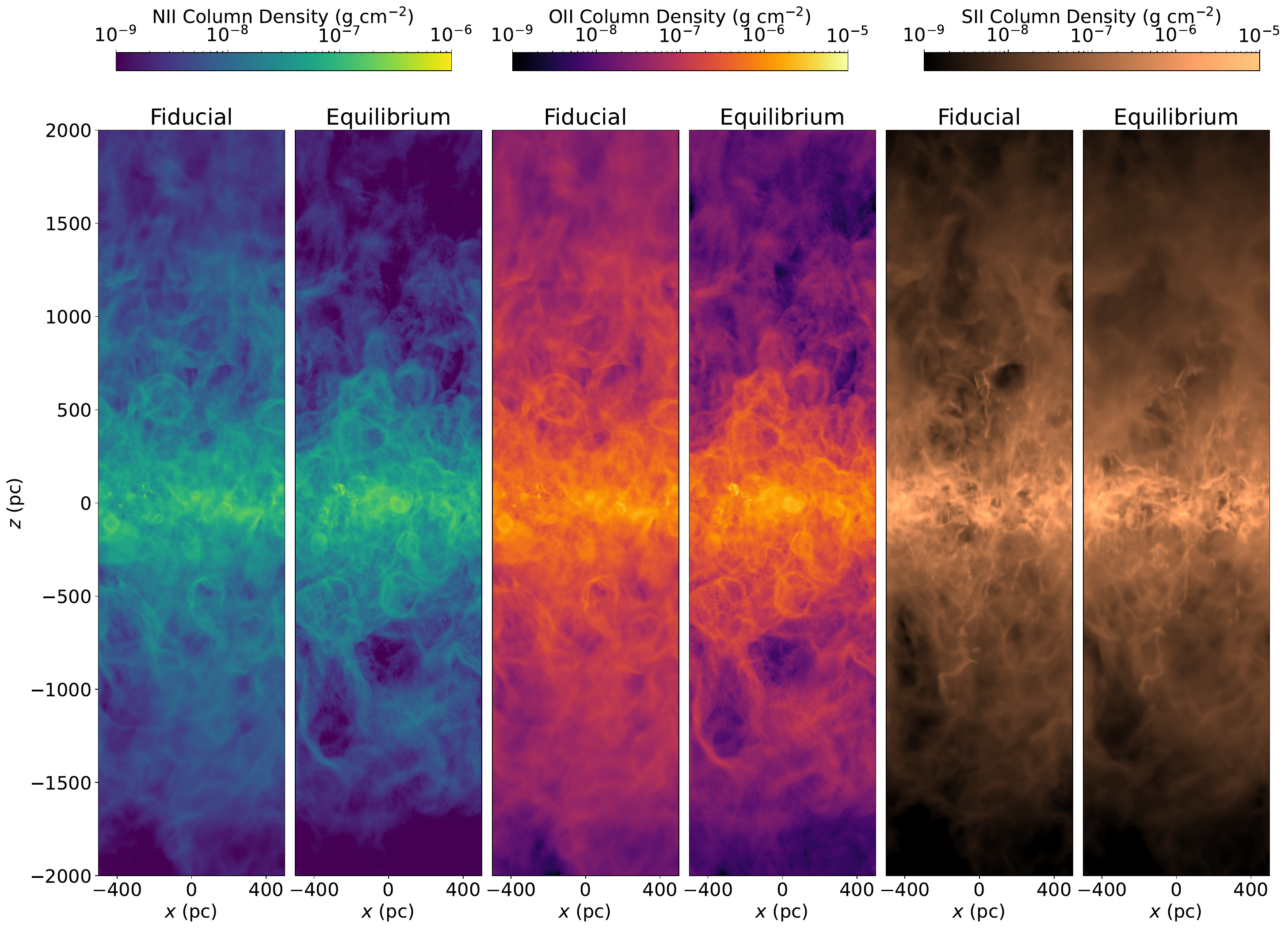}
    \caption{Comparison of selected ion structures after 25~Myr of evolution of the full radiation hydrodynamics code. Left panels of each pair show the results of the full time dependent simulation, with equilibrium ionization structures shown in the right panel of each pair. Top row of structures shows higher ionization energy ions. Bottom row shows the lower ionization energy ions of NII and OII, alongside SII which is the ground state of sulphur in our simulations. In general higher ionization energy ions shows a larger difference in extent, the low ionization energy ions show a moderate change from equilibrium to time dependent, and the ground state of SII shows minimal difference.}
    \label{struct} 
\end{figure*}

\subsubsection{Atomic Data}

The atomic data in our simulations are the same as in \citet{cmacionize} for ionization cross sections (from \citet{verner95}) and charge transfer rates. The radiative recombination rates, dielectronic recombination rates, and collisional ionization rates are from the CHIANTI database \citep{delzanna21}. This is a departure from the recombination and collisional ionization rates used in \citet{vandenbroucke18}.

\subsection{Cooling Rates}

The simulations of \citet{mccallum24} employed cooling rates from the tables of \citet{derijke13}. Whilst a functional approximation for the conclusions of \citet{mccallum24}, this method does not parametrize the ionization state of the gas, and hence precise losses to collisionally excited emission lines are not calculated. It is known that the calculation of emission line intensity is sensitive to the temperature of the gas, so we now explicitly calculate the line cooling between every hydrodynamical timestep using the current temperature and metal ionization state of each cell as in \citet{wood04} and \citet{cmacionize}. The collisionally excited emission line cooling is expected to be dominant only in the warm temperature regime. For this reason, the original cooling curve of \citet{derijke13} is implemented when the temperature is below 3000 K or above 50,000 K. A cooling floor of 100~K is implemented, whereby gas is not able to cool below this temperature.

\subsection{Overview of Simulations}

Four simulations are run from the same initial conditions. The \emph{fiducial} run includes the full time dependent ionization state calculation and cooling rate calculation from the collisionally excited emission lines. The \emph{equilibrium} run includes the full cooling calculation but will calculate the ionization state of each ion using the assumption of ionization equilibrium. The \emph{table cooling} run calculates the ionization states in a time-dependent manner, but retrieves cooling rates from the tables of \citet{derijke13} (as in \citet{mccallum24}). The \emph{HOLMES} run includes the population of HOLMES sources from the HOLMESLOW simulation of \citet{mccallum24} (1000 sources, each with ionizing luminosity $5\times10^{45} \rm{s}^{-1}$, and scale height of 700~pc), time-dependent ionization and full temperature calculation. These parameters for the population of HOLMES sources was selected as having given the best match to observed DIG densities and scale heights in the hydrogen-only simulations of \citet{mccallum24}.

\section{Benchmarking time-dependent ionization}
\label{benchmarking}


What have become a standard set of benchmarks for the testing of ionization codes are those from the 1995 Lexington workshop \citep{ferland95}, referred to here as the Lexington benchmarks and described in detail in \citet{pequignot01}.

Of the set of benchmarks described in \citet{pequignot01}, we focus on HII20 and HII40. These have a uniform density of $100~{\rm cm}^{-3}$ (zero density inside a radius of 1~pc) around a central source with black body spectra for temperatures of 20kK and 40kK respectively. The resulting ionization and temperature structure and intensities of various emission lines can then be compared to other ionization codes.

\citet{cmacionize} tested CMacIonize adopting ionization equilibrium, and confirmed a close match with other codes, such as the widely used 1D ionization equilibrium code \texttt{CLOUDY}. Here we replicate these benchmarks in CMacIonize, but use the time-dependent scheme to resolve the development of the ionized regions over a 3000~yr period by which time the ionization and temperature structure has reached a steady state.

Analogous to the time-resolved development of the ionized region, we now must also track the temperature evolution explicitly throughout the simulation. Rather than iterating to an equilibrium temperature as in \citet{cmacionize}, we use the cooling rates determined from the metal ionization states, and explicitly evolve the temperature of each cell, balancing cooling with photoionization heating of hydrogen and helium at each timestep.

Alongside runs explicitly calculating the cooling within each cell to determine the temperature evolution, we run two further tests of our time-dependent Lexington benchmarks in which the cooling curves from \citet{derijke13} will define the cooling rate. As the underlying simulations from \citet{mccallum24} do not explicitly track the ionization state of each metal, the tabulated curves of \citet{derijke13} were used to calculate cooling rates. These tests will investigate differences between tabulated cooling and explicitly-tracked collisionally excited line cooling.

\subsection{HII20 Benchmark}

The time-dependent simulations of the HII20 benchmark in CMacIonize give an ionization state at the end of the simulation period which matches closely with the equilibrium results obtained from \texttt{CLOUDY} for every ion included in the benchmark (figure \ref{lex20}).

The temperature profiles obtained using tabulated cooling rates from \citet{derijke13} exhibit temperatures that are higher and almost uniform with radius. This is clearly very different from  those calculated with explicit cooling based on the ionization states of the metals (figure \ref{lex20table}).

When calculating the emission line intensities using the full cooling model, the values match those from \texttt{CLOUDY} within 26\% in all but the [OI] 6300~\AA\ line. As discussed in \citet{wood04}, this is likely due to poor photon sampling at the outer edge of the ionized volume, which is the region associated with emission from neutral oxygen. 

The emission line intensities derived from the tabulated cooling runs do not reproduce the \texttt{CLOUDY} values. The resulting line intensities from each Lexington benchmark run can be found in the appendix.

\begin{figure*}
    \centering
    \includegraphics[width=0.7\textwidth]{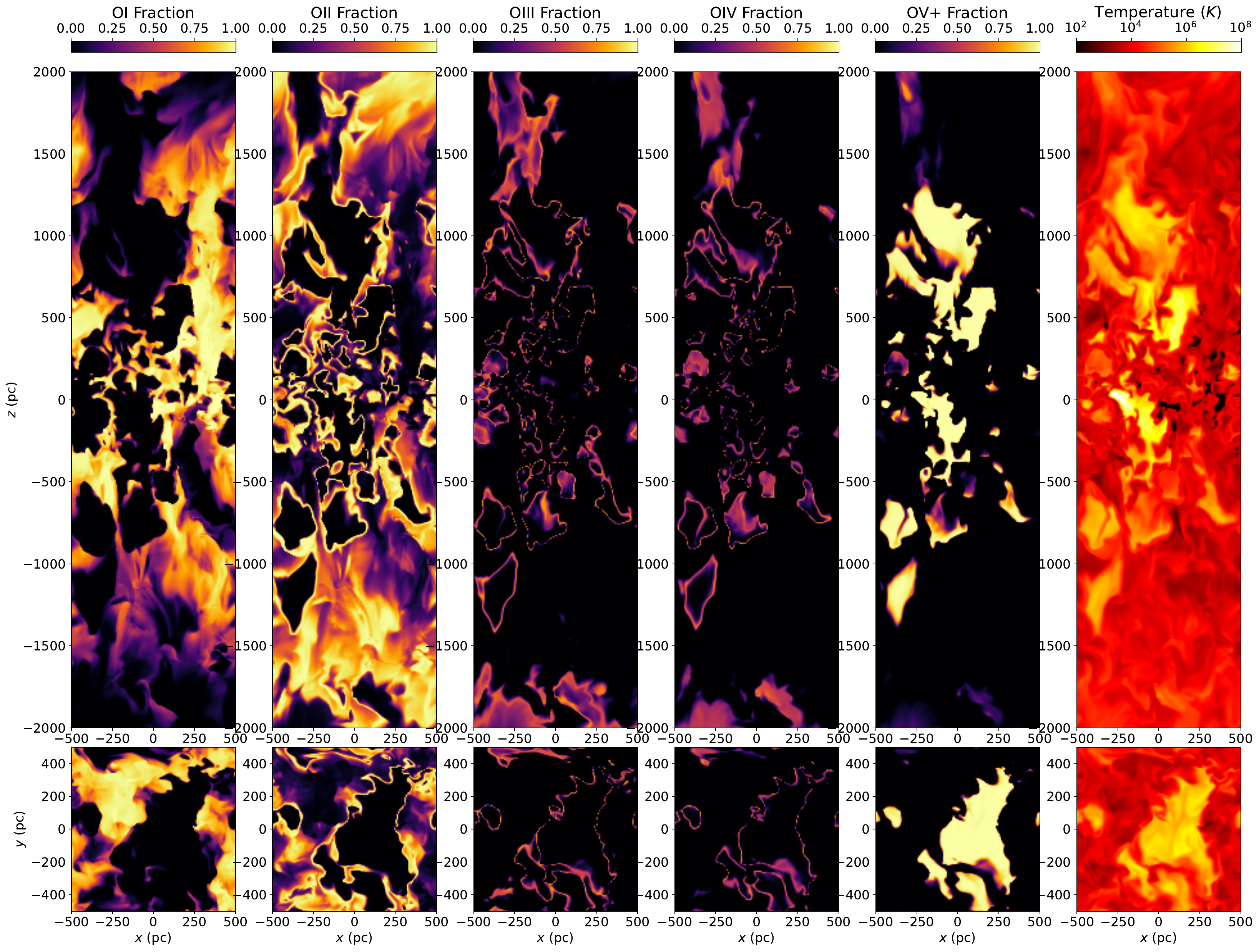}
    \caption{Slices of oxygen ion fractions in our \emph{fiducial} simulation after 25~Myr of evolution from the initial conditions. Top row shows a central slice through the X-Z plane, with the bottom row showing a slice through the midplane (in X-Y). The right hand column shows the temperature of each cell shown in the other slices. An 'onion skin' model around supernova bubbles appears to hold true, whereby lower energy ions are found in successive shells around the hot bubbles. Ions sensitive to OB star photoionization show further structure at altitude. The second from the right column shows all ions of OV and above.}
    \label{oxygenstructure} 
\end{figure*}

\begin{figure}
    \centering
    \includegraphics[width=0.9\columnwidth]{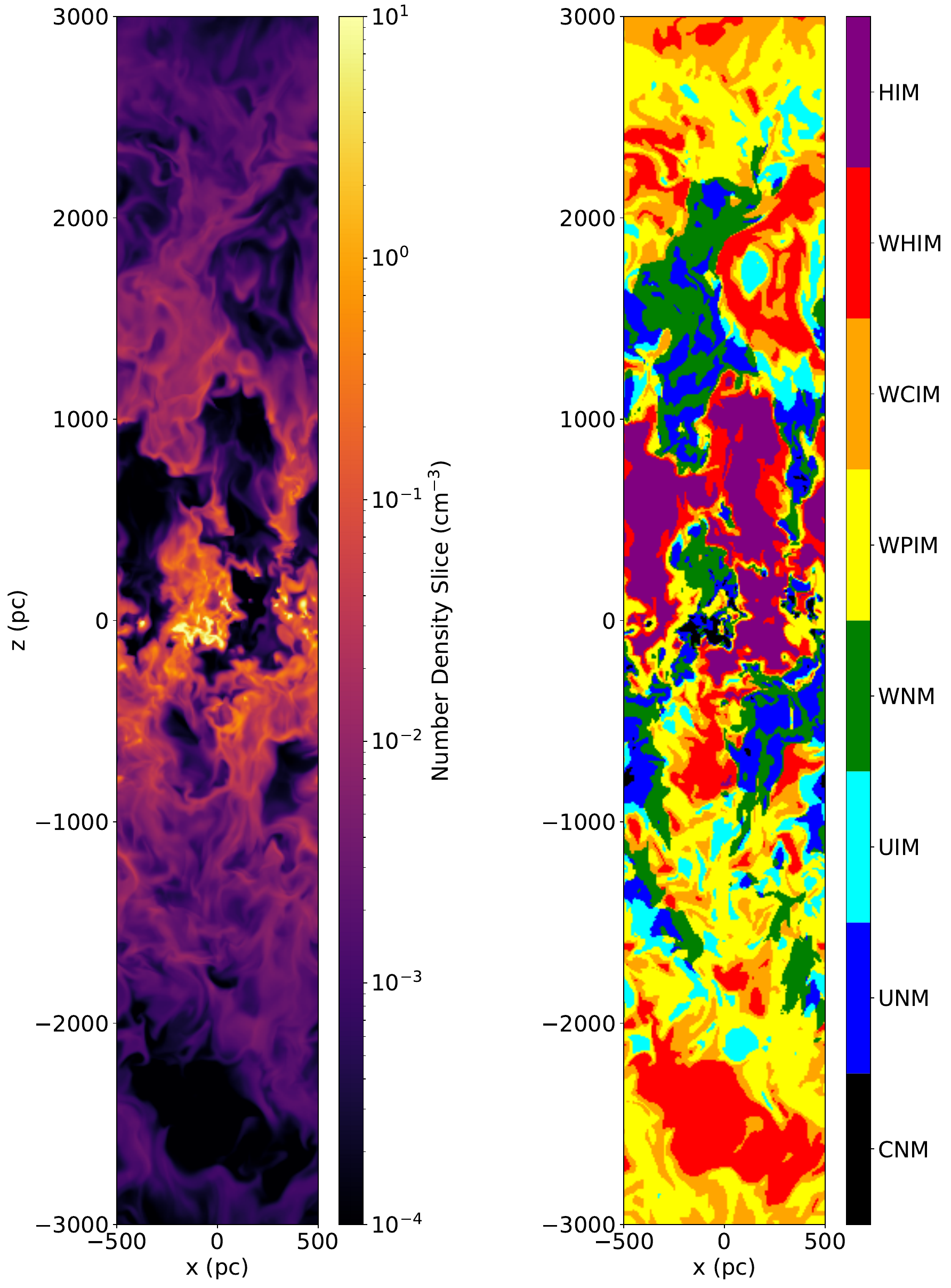}
    \caption{Right panel shows gas phases in a central slice of our \emph{fiducial} simulation after 25~Myr of evolution. Gas phase definitions are from \citet{kim23}. The left panel shows the number density in the equivalent slice.}
    \label{phasesfromkim} 
\end{figure}

\begin{figure}
    \centering
    \includegraphics[width=0.7\columnwidth]{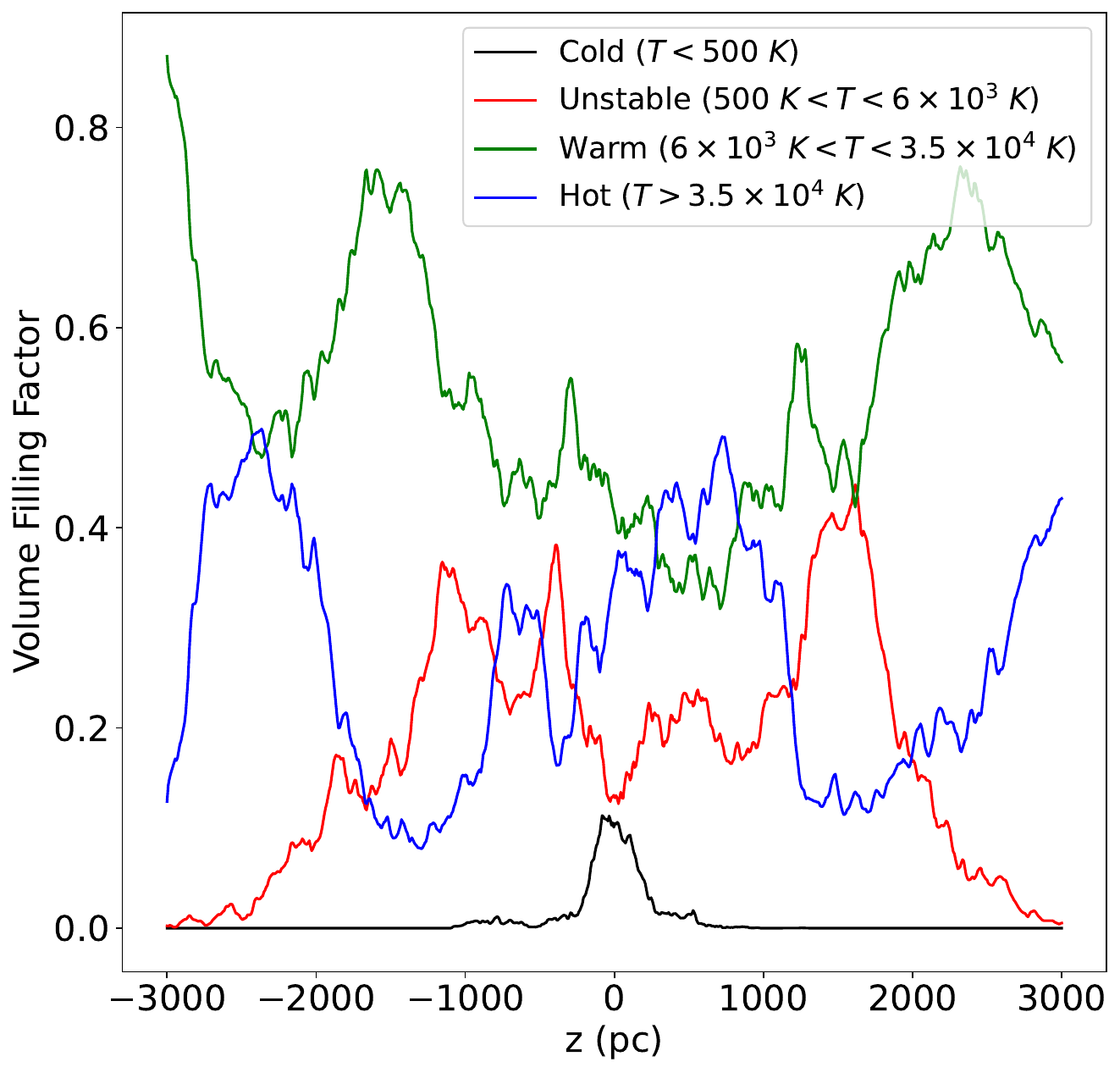}
    \caption{Height versus volume filling factors of gas in 4 main temperatures phases in our \emph{fiducial} simulation after 25~Myr of evolution.}
    \label{vffs} 
\end{figure}

\begin{figure}
    \centering
    \includegraphics[width=1.0\columnwidth]{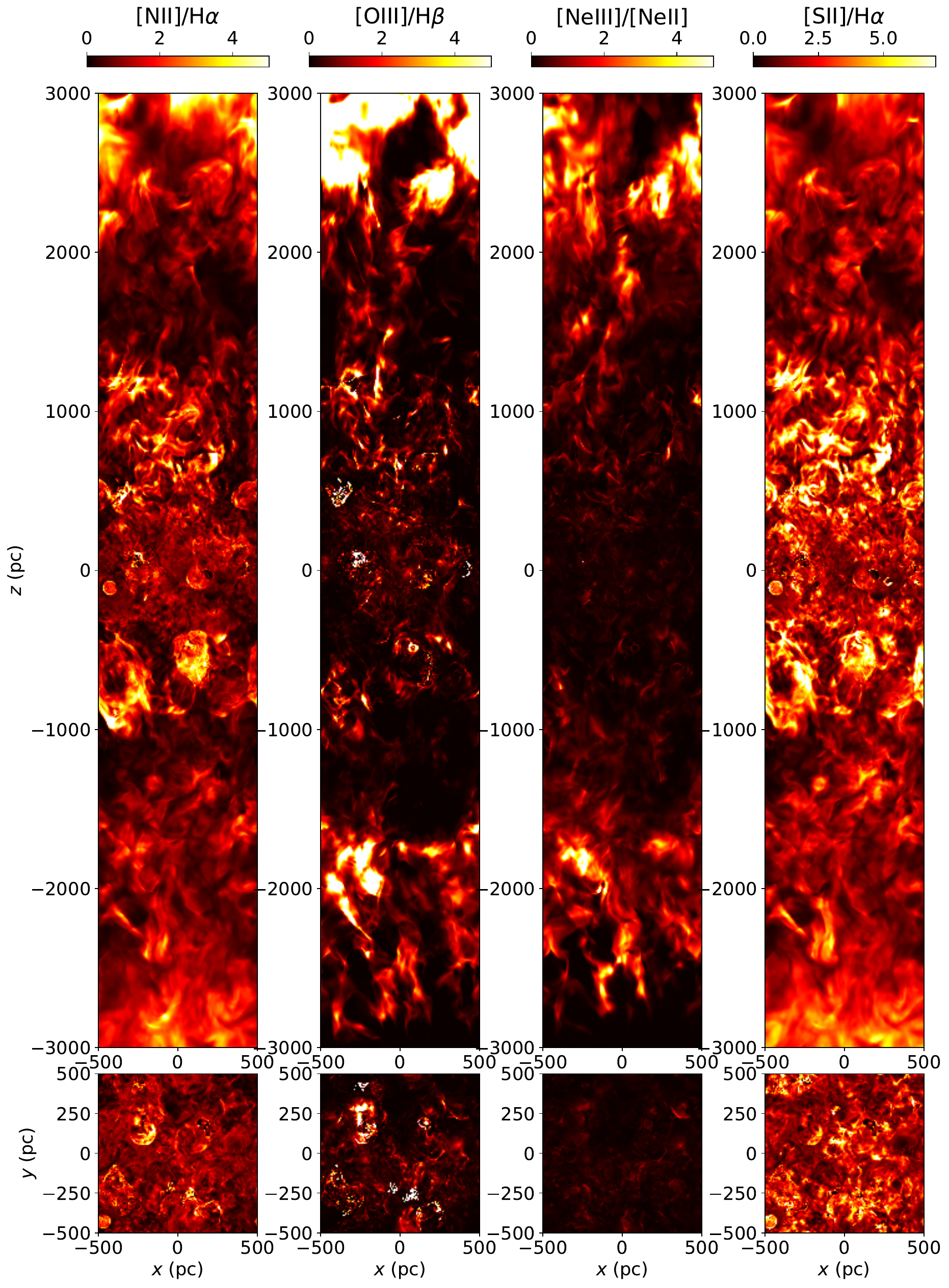}
    \caption{Maps of edge on line ratios in our \emph{fiducial} simulation after 25~Myr of evoution from the initial conditions. General rising trends are seen with height in all line ratios, with the exception of midplane values of [NII]/H$\alpha$ and [SII]/H$\alpha$ which are elevated.}
    \label{ratiostruct} 
\end{figure}

\begin{figure*}
    \centering
    \includegraphics[width=0.9\textwidth]{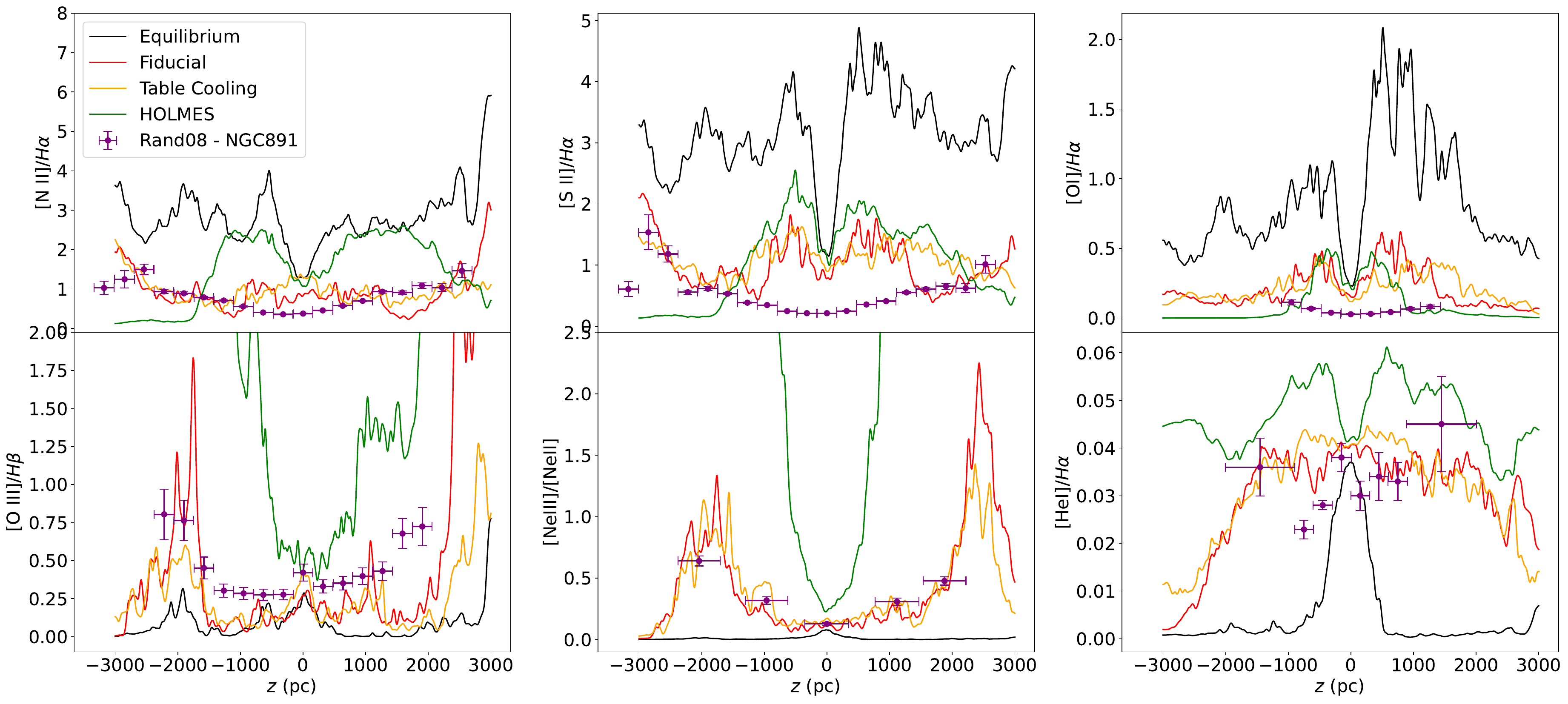}
    \caption{Selected line ratios versus height above midplane for our simulations after 25~Myr of evolution from the initial conditions. Also shown in purple dots is the vertically resolved observational data from NGC 891 of \citet{rand08}. Horizontal error bars show the z-range over which the observational data was averaged. Simulated line ratios are a median value for each value of z.}
    \label{lineratios} 
\end{figure*}

\begin{figure}
    \centering
    \includegraphics[width=0.9\columnwidth]{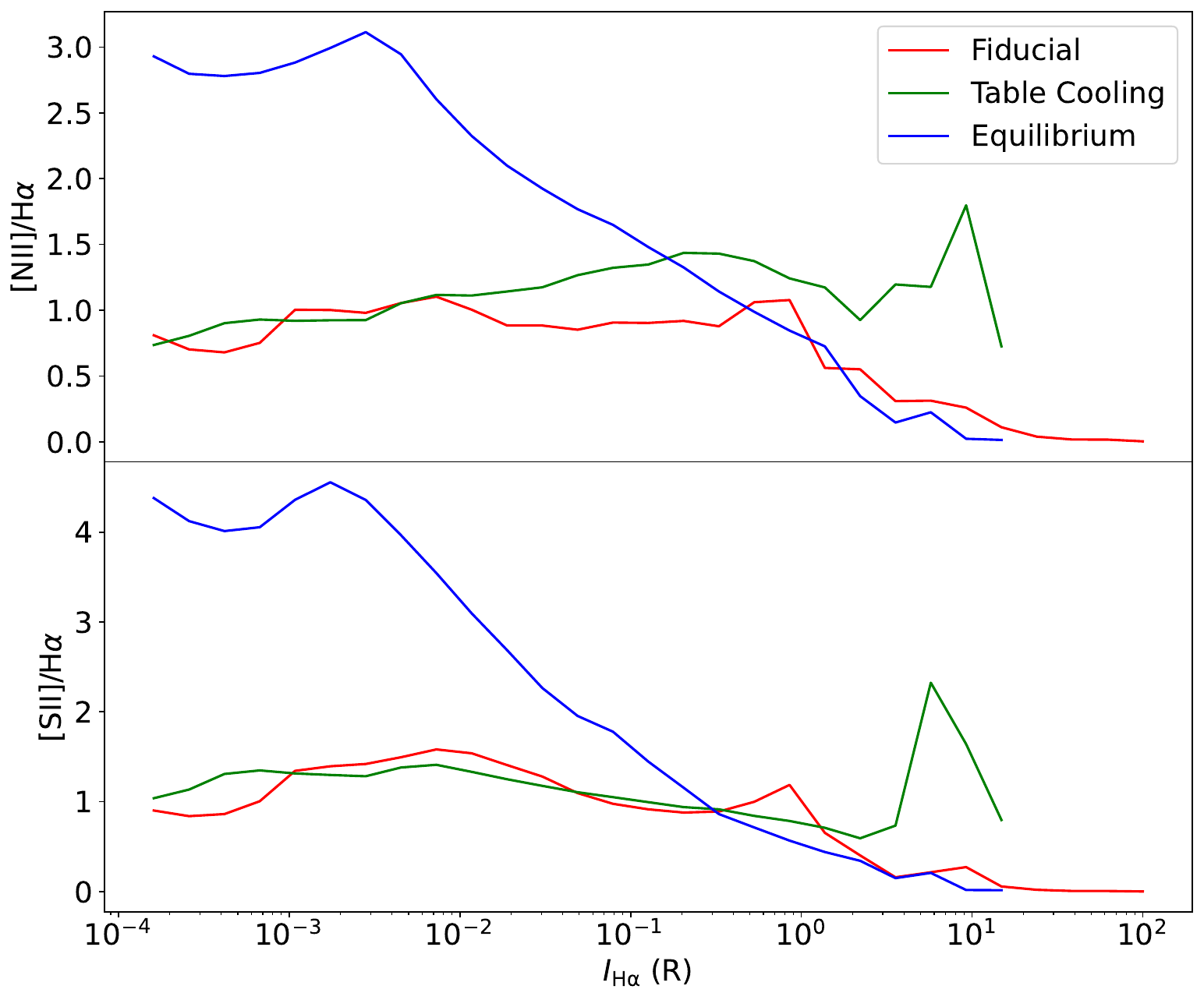}
    \caption{Mean line ratios of [NII]/H$\alpha$ and [SII]/H$\alpha$ as a function of H$\alpha$ surface brightness from edge on synthetic emission maps. \emph{Equilibrium} shows a steep rise in each line ratio with decreasing H$\alpha$, whereas both \emph{table cooling} and \emph{fiducial} level off at values of around 1.0 and 1.2 respectively. \emph{Table cooling} shows a spike in both ratios towards the high H$\alpha$ values, likely due to nonphysically hot gas due to underestimated cooling in denser midplane photoionized gas.}
    \label{ratiovhalpha} 
\end{figure}

\begin{figure}
    \centering
    \includegraphics[width=0.9\columnwidth]{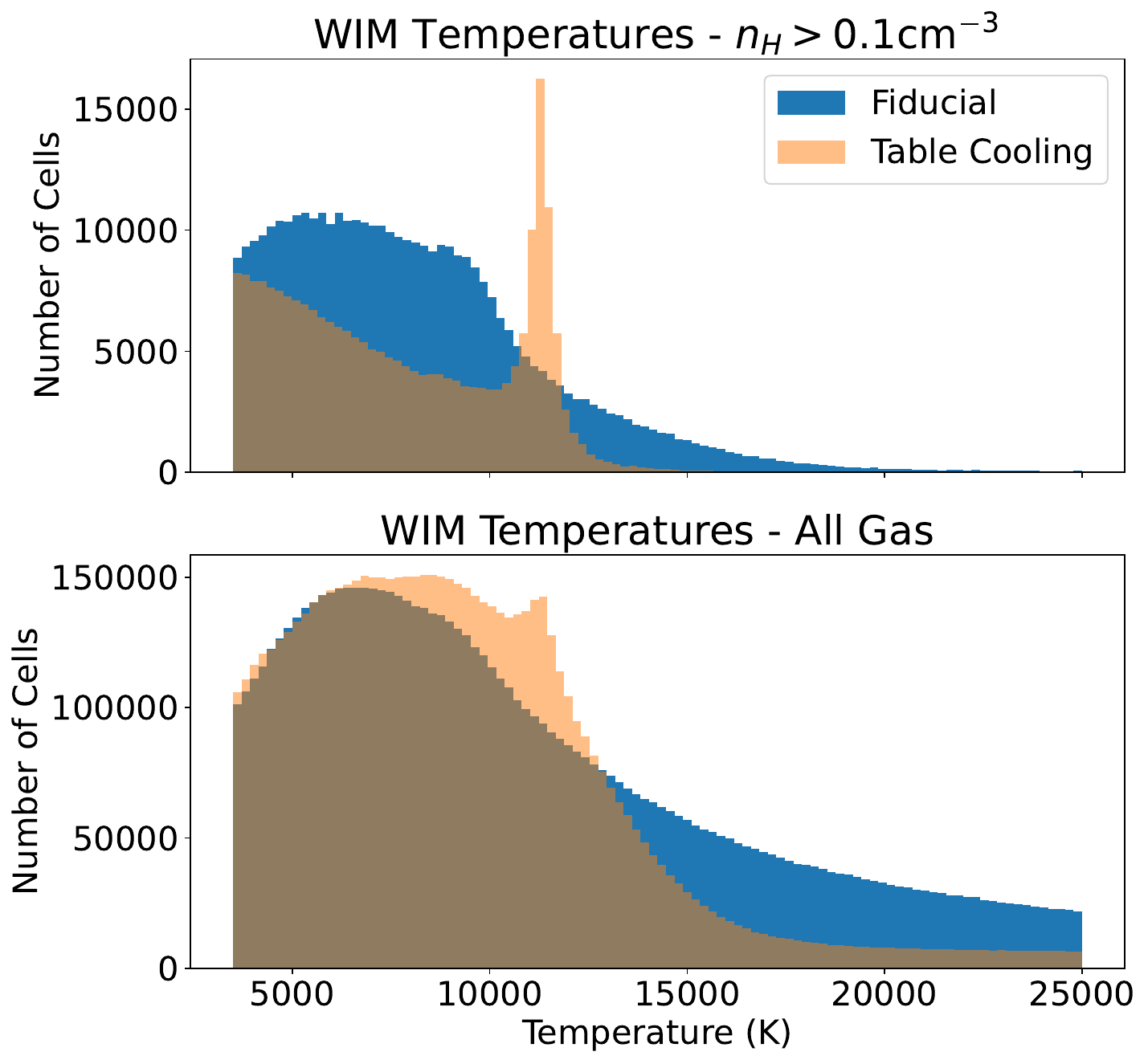}
    \caption{Histograms of gas temperatures between 3500 and 25000 K. Top panel is only showing gas in the denser regime of $n_{H} > 0.1 \rm{cm}^{-3}$, bottom panel shows all gas. \emph{Table cooling} and \emph{fiducial} are seen to diverge in temperature structure in the gas which is $> 0.1 \rm{cm}^{-3}$, but globally the gas temperature structure is broadly similar.}
    \label{temphisto} 
\end{figure}

\subsection{HII40 Benchmark}

For the HII40 benchmark, the time-dependent simulations also show that the ionization state at the end of the simulation is in agreement with the equilibrium results from \texttt{CLOUDY} (figure \ref{lex40}). The temperature profiles for the HII40 benchmark reveal similar patterns to the HII20 benchmark. The runs using tabulated cooling rates produce higher temperatures and a more uniform temperature structure compared to the explicit cooling calculations (figure \ref{lex40table}).

For the emission line intensities, the full cooling model produces values that are within 26\% of the \texttt{CLOUDY} results for most emission lines. This again excludes the [OI] 6300~\AA\ line, for which \texttt{CMacIonize} overestimates this emissivity due to poor photon sampling at the edge of the ionized region. We also find moderate overestimations (26-53\%) for [OIII] 88~$\mu m$, [NeIII] 3869~\AA\ and [NeIII] 15.5~$\mu$m.

The tabulated cooling runs show significant deviations from the \texttt{CLOUDY} line intensities, again indicating that tabulated cooling rates are not appropriate when simulations require accurate temperatures and line emissivities from photoionized gas.

\section{Full Radiation-Hydrodynamics Results}

\subsection{3D Structure}

The result from \citet{mccallum24} that the time-dependent calculation of the ionization state of hydrogen is crucial for modelling the low density, high altitude DIG is replicated in these shorter timescale simulations including metals, with the \emph{equilibrium} and \emph{fiducial} runs showing differences in hydrogen ionization structure. 
This difference in hydrogen ionization structure is also seen in the structure of ions with similar ionization energy to hydrogen (eg. \ion{N}{II} , \ion{O}{II} ). Figure \ref{struct} shows the general deficit in the production of these ionization fractions compared to the time-dependent run.

The biggest difference in ionization structure between the \emph{equilibrium} and \emph{fiducial} runs is in the production (and persistence) of ions which high ionization potentials (e.g., \ion{Ne}{III} , ionization energy 41.0~eV). Photons with energies above 41~eV are rare with 40kK OB star spectra, and under the assumption of ionization equilibrium very little \ion{Ne}{III}  is created. \ion{Ne}{III}  is only seen in the midplane where the flux of high energy photons is enough to support an equilibrium population in that ion. However with the time-dependent calculation, the slow recombination rates at low densities from \ion{Ne}{III}  back to \ion{Ne}{II}  causes an overionization (compared to equilibrium) of \ion{Ne}{III}  at high altitudes. The strength of this overionization is dependent on the gas density and is hence seen to be a function of height above the midplane.

In our fiducial simulation, we find that the collisionally ionized metals are structured in shells around hot bubbles, with the highest ionization states existing in the innermost regions of the bubble, and successive shells of each lower ionization state. These structures are shown in figure~\ref{oxygenstructure}, displaying single pixel-wide slices of the ionization fraction of successive oxygen ions.

\begin{figure*}
    \centering
    \includegraphics[width=0.7\textwidth]{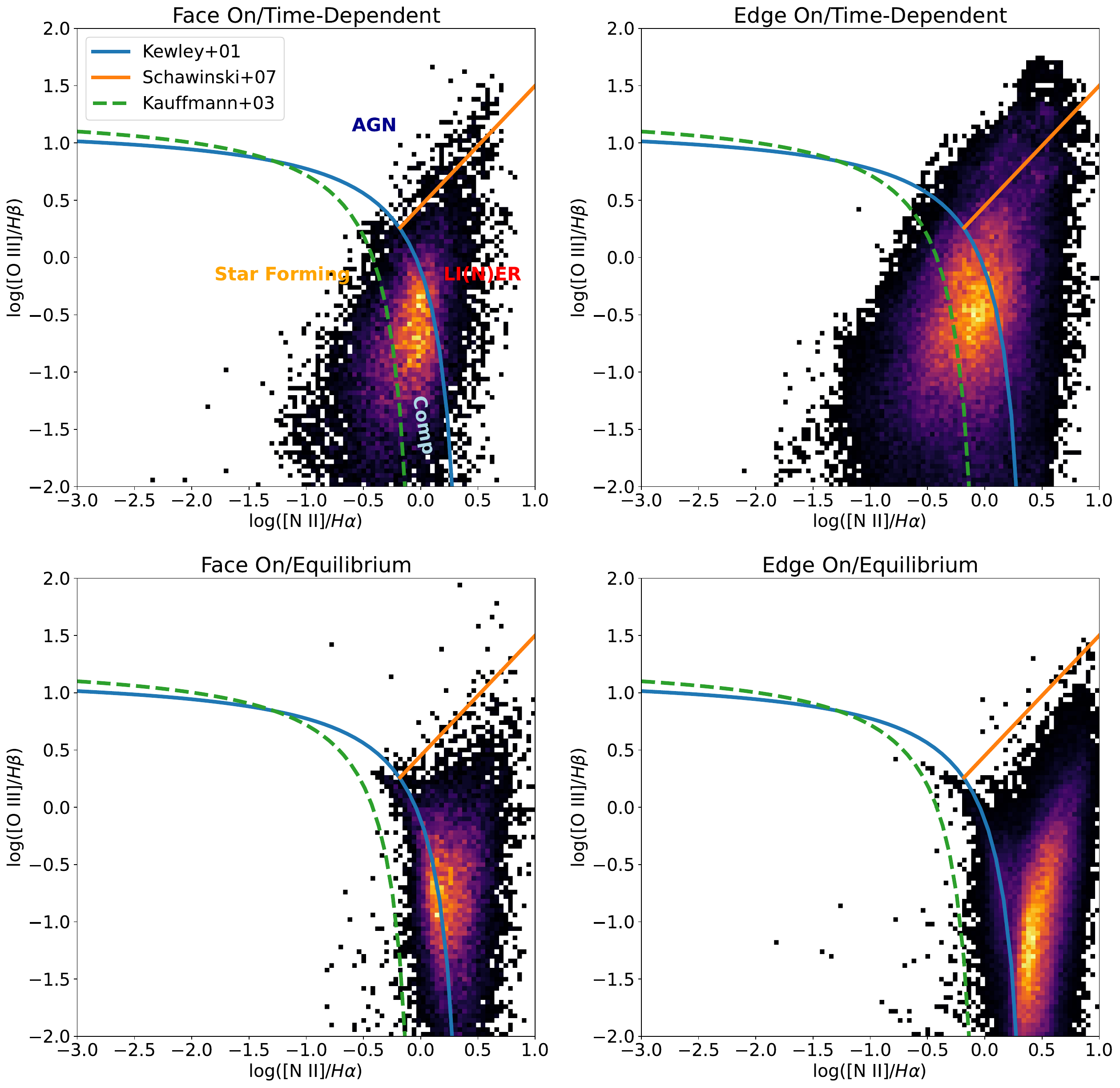}
    \caption{[NII]/[OIII] BPT diagram for both the \emph{fiducial} simulation (top row) and \emph{equilibrium} simulation (bottom row) after 25~Myr of evolution from the initial conditions. The green, blue and orange lines separate the labelled regions of the BPT diagram by ionization mechanism (\citet{kewley01}; \citet{kauffmann03}; \citet{shcawinski07}). The left column shows BPT diagrams made from face on images, and the right column is for edge on images. Z-scale in these BPT diagrams is number of pixels, meaning low emission measure, low density gas contributes equivalently to the higher density ionized gas.}
    \label{bptpanels}
\end{figure*}

Figure~\ref{phasesfromkim} shows the phase structure of the central pixel-wide slice from our \emph{fiducial} simulation after 25~Myr of evolution. The phase definitions are from \citet{kim23} and include the following media; Cold Neutral Medium (CNM), Unstable Neutral Medium (UNM), Unstable Ionized Medium (UIM), Warm Neutral Medium (WNM), Warm Photoionized Medium (WPIM), Warm Collisionally Ionized Medium (WCIM), Warm Hot Ionized Medium (WHIM), Hot Ionized Medium (HIM). The volume filling factor (VFF) of 4 temperatures regimes is also shown in figure~\ref{vffs}. Our simulations produce a significant volume of unstable neutral and unstable ionized medium, with the filling factor of the unstable gas being higher at high altitudes (lower densities), peaking at an unstable gas VFF of 0.4 between 1 and 2~kpc of height. This is to be expected where the lower density gas takes longer to transition between stable temperatures due to the density dependence of the cooling rates.

\subsection{Synthetic Line Ratios}

We produce edge-on emission line maps for every snapshot of every simulation. Edge-on maps are produced of the following emission lines; [NII] 6584~\AA, [SII] 6716~\AA, [OI] 6300~\AA, [OIII] 5007~\AA, [NeIII] 15~$\mu m$, [NeII] 12~$\mu m$, [HeI] 5876~\AA, H$\alpha$ and H$\beta$. These lines will be used to compare to the spectroscopic data from the Milky Way analogue NGC~891 presented in \citet{rand08}, taken at various heights above and below the midplane at a radius of $R=100"$, or 4~kpc assuming a distance of 8.4~Mpc.

The edge-on emission line ratio maps for our fiducial simulation are shown in figure~\ref{ratiostruct}, where a general increase in each line ratio with height is visible, other than in the midplane for [NII]/H$\alpha$ and [SII]/H$\alpha$.

The \emph{fiducial} and \emph{equilibrium} run results after 25~Myr of evolution are shown in figure \ref{lineratios}. We quantify the quality of fit to observed data by accumulating the line intensities across the same vertical bins as \citet{rand08} and calculating the normalized root mean squared error (NRMSE). Each error is first normalized to the value of the observed data, and the mean square of each normalized error is computed (and square rooted) to calculate the NRMSE. The deviations for our \emph{fiducial} model after 25~Myr of evolution are displayed in table~\ref{fitquality}, both for the entire simulation box, and separately for high and low altitude gas (defined as above and below 1~kpc of height.)

\begin{table*}
	\centering
	\caption{NRMSE between \emph{fiducial} simulation at 25~Myr and data from \citet{rand08}.}
	\label{fitquality}
	\begin{tabular}{lccc} 
		\hline
		Line Ratio & NRMSE (full box) & NRMSE ($|z|< 1~kpc$) & NRMSE ($|z|>1~kpc$)\\
		\hline
		$[\rm{NII}]$/H$\alpha$ & 0.73 & 1.04 & 0.37\\
		$[\rm{OIII}]$/H$\beta$ & 0.42 & 0.28 & 0.53\\
		$[\rm{SII}]$/H$\alpha$ & 2.14 & 3.39 & 0.46\\
            $[\rm{NeIII}]$/$[\rm{NeII}]$ & 0.36 & 0.44 & 0.29 \\
            $[\rm{HeI}]$/H$\alpha$ & 0.29 & 0.33 & 0.12 \\
            $[\rm{OI}]$/H$\alpha$ & 7.55 & 7.96 & 3.56 \\
		\hline
	\end{tabular}
\end{table*}

Our \emph{fiducial} simulation produces line ratios for the high altitude, low density gas which are comparable to those observed in NGC 891 for [NII]/H$\alpha$, [SII]/H$\alpha$, [OIII]/H$\beta$, and [NeIII]/[NeII]. Midplane line ratios are matched for [OIII]/H$\beta$, [NeIII]/[NeII] and [HeI]/H$\alpha$.

The line ratios of [NII]/H$\alpha$ and [SII]/H$\alpha$ at $|z| < 1 \rm{kpc}$ from our simulations are are larger than observed in NGC~891 (with NRMSE values > 1), and [OI]/H$\alpha$ is overpredicted, as is seen in the benchmarking.

The equilibrium line ratios are very different from the observed NCG~891 values, with larger than observed for with low ionization lines, and much smaller for higher ionization potential ions such as [OIII] and [NeIII]. The [NeIII]/[NeII] line ratio structure with height displays the largest difference from time-dependent to equilibrium, with almost no \ion{Ne}{III}  existing under the assumption of ionization equilibrium and the 40~kK spectrum used. 

The vertically resolved emission line ratio maps from the \emph{table cooling} run are also displayed in figure~\ref{lineratios}. We find that in the full radiation hydrodynamics simulations including feedback from both photoionization and supernovae, the tabulated cooling rates replicate the emission line structures at high altitudes from low density gas very well. This is despite the \emph{table cooling} method failing to appropriately model the expected results in the Lexington benchmark runs. We do find however that temperature sensitive line ratios of [NII]/H$\alpha$ and [SII]/H$\alpha$ are overestimated in denser regions of the midplane DIG where photoionization is the most dominant form of ionization. Whilst not easily apparent in figure~\ref{lineratios} due to the midplane representing only a small extent in $|z|$, figure~\ref{ratiovhalpha} shows the mean line ratio value in each H$\alpha$ surface brightness bin for edge-on synthetic images. A large overestimation of of these line ratios at high values of H$\alpha$ surface brightness (a proxy for DIG density) can be seen for the \emph{table cooling} run. To further investigate the deficit in cooling rates in $n_{H}$ > 0.1 $\rm{cm}^{-3}$ gas, figure~\ref{temphisto} shows the distribution of DIG temperatures for the full simulations, and for only the denser gas. It is seen that the high temperature gas excess in the \emph{table cooling} run is seen to be mostly constrained to the denser gas.

Our \emph{HOLMES} simulation results in line ratios which are very different from those observed in NGC~891, in particular for high energy ions such as \ion{O}{III}  and \ion{Ne}{III} , which are found in much higher abundances. Values of high altitude [NII]H/$\alpha$ are either found to be higher than observed due to overheating of the DIG, or much lower due to the further ionization of \ion{N}{II}  to \ion{N}{III} . These results are also displayed in figure~\ref{lineratios}.

\subsection{Diagnostic Emission Line Diagrams}

\subsubsection{BPT Diagrams}

From our \emph{fiducial} simulation we produce diagnostic diagrams from projected 2D face-on images of specific emission lines. We focus on two diagnostic diagrams; the BPT diagrams (as first described by \citet{baldwin81}) of [NII]/$\rm{H}\alpha$ to [OIII]/$\rm{H}\beta$ and [NII]/$\rm{H}\alpha$ to [SII]/$\rm{H}\alpha$. The former of which is used to diagnose the ionization mechanism of the nebular gas, and the latter being utilised to estimate the gas temperature and ionization state. 

The [NII]/$\rm{H}\alpha$ versus [OIII]/$\rm{H}\beta$ BPT diagram distinguishes 3 regions of differing ionization mechanism; photoionization due to active star formation, active galactic nuclei, and low-ionization (nuclear) emission line regions (LI(N)ERs). LI(N)ER gas was introduced by \citet{heckman80}, with the ionization mechanism of this region being less well constrained.

Figure~\ref{bptpanels} shows the results of the [NII]/[OIII] BPT diagram, in both the \emph{fiducial} and \emph{equilibrium} runs, for edge-on and face-on viewing. These diagrams display a density of pixels in each region of the diagram. The \citet{kewley01} {\it maximum starburst line} is also shown as a theoretical limit on gas photoionized by massive stars. In the \emph{fiducial} run, most pixels fall below the starburst line in the face-on view, with more pixels located above the starburst line for edge-on viewing. This pattern is due to greater contributions of the low-density (low intensity) high altitude gas, and consistent with the rising [NII]/H$\alpha$ and [OIII]/H$\beta$ line ratios with altitude. The low density gas is therefore more likely to be classified as LI(N)ER gas than starburst ionized gas, with the AGN region of the diagram sparsely populated in all of our results.

The simulation where the gas is in ionization equilibrium, with heightened [NII]/H$\alpha$ (as seen in figure~\ref{lineratios}), has the LI(N)ER section of the diagram populated, with the face-on BPTs varying less from time-dependent to equilibrium than the edge-on diagrams. This is again consistent with the result that time-dependent effects are primarily a feature of the low-density high-altitude gas, which due to its low surface brightness is less prominent in face-on images.

\subsubsection{Characteristics of gas on the diagnostic diagrams}

\begin{figure*}
    \centering
    \includegraphics[width=0.8\textwidth]{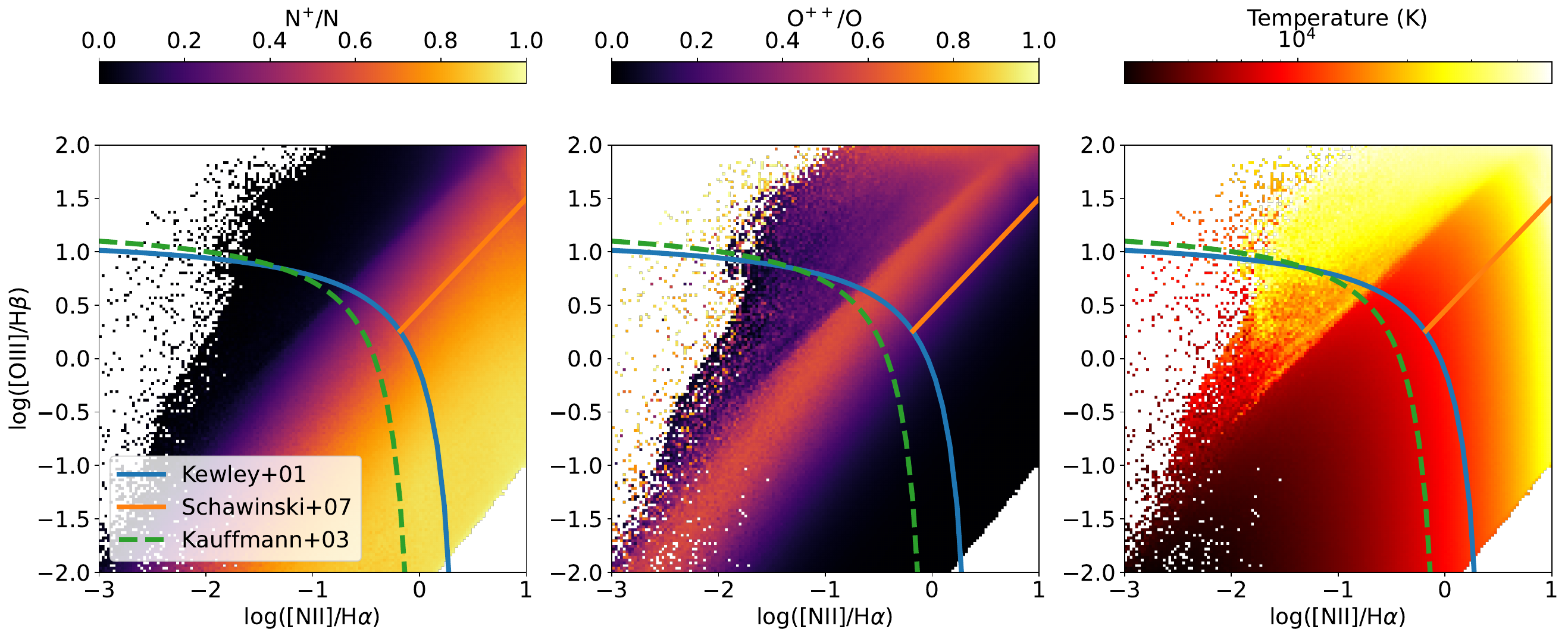}
    \caption{Ionization state and temperature gas between 3000~K < T < 50,000~K as a function of location on the [NII]/[OIII] BPT diagram. Results are from the \emph{fiducial} simulation at a time of 25~Myr.}
    \label{o3n2fig}
\end{figure*}

\begin{figure*}
    \centering
    \includegraphics[width=0.8\textwidth]{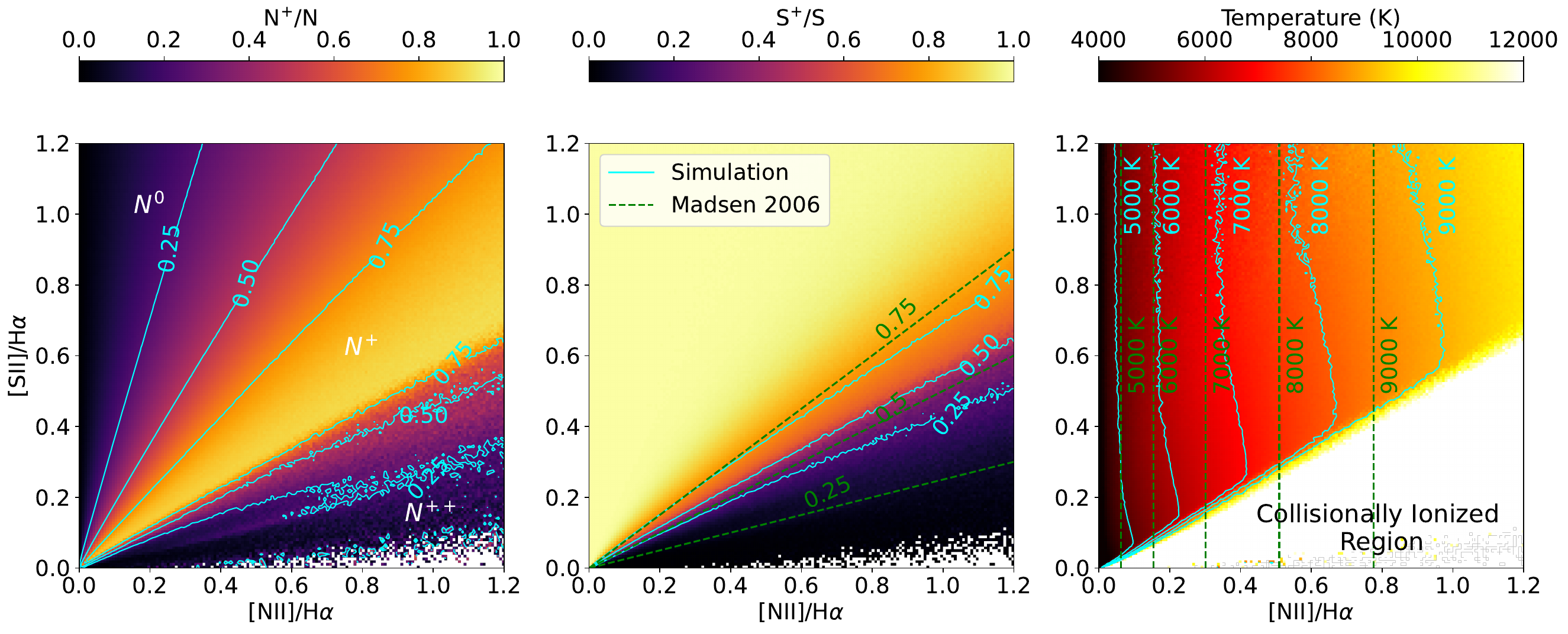}
    \caption{Ionization state and temperature of the gas between 3000~K < T < 50,000~K as a function of location on the [NII]/[SII] diagnostic diagram. Results are from the \emph{fiducial} simulation at a time of 25~Myr. Cyan lines show contours from the simulation data, and green dashed lines show expected contours from observational theory described in \citet{madsen06}.}
    \label{s2n2fig}
\end{figure*}

To better understand the physical conditions which lead to emission in certain regions of the two diagnostic BPT diagrams, we bin all voxels in our \emph{fiducial} simulation at 25~Myr by location on the BPT diagram, producing maps of ionization state and temperature. Figure~\ref{o3n2fig} shows the results for the [OIII]/[NII] BPT diagram, and figure~\ref{s2n2fig} shows the diagnostic diagram of [SII]/[NII].

In the [OIII]/[NII] BPT diagram, there is a general trend of increasing temperature with increasing [NII]/H$\alpha$. The nitrogen ionization state of the gas increases uniformly across the map from the lower right quadrant to the upper left, with upper left being deficient in \ion{N}{II}  due to further ionization to \ion{N}{III} . A similar trend is seen in \ion{O}{III} , with the lower right quadrant being low in \ion{O}{III}  due to \ion{O}{II}  and \ion{O}{I}  being the dominant ions, through a region of high \ion{O}{III}  towards the upper left quadrant, where warmer/hotter, more collisionally dominated gas resides in states further ionized above \ion{O}{III} .

\citet{haffner09} provides a description of the use of the diagnostic diagram of [NII]/H$\alpha$ versus [SII]/H$\alpha$. One expects temperature to increase from left to right with rising [NII]/H$\alpha$ and sulphur ionization state to change from doubly to singly ionized moving from the lower right quadrant to upper left. Figure~\ref{s2n2fig} displays both of these expected trends. It is also seen that there is a steep jump in the mean temperature moving into the lower right quadrant of the diagram, from typical DIG temperatures of approximately $10^{4}$~K, to nearer $4-5\times10^{4}$~K. This region is labelled as the 'collisionally ionized region', as it contains mostly gas above typical DIG temperatures, with low \ion{N}{II}  due to an excess of \ion{N}{III} .

Also compared in figure~\ref{s2n2fig} are the expected contours for temperature and sulphur ionization state \citep{madsen06} against those derived from the \emph{fiducial} simulation. The temperature contours are seen to match closely to expected, with the exception of the collisionally ionized 'wedge' in the lower right section of the diagram. This type of gas is not considered by \citet{madsen06}, in which collisionally ionized gas is further ionized from \ion{N}{II}  to \ion{N}{III} . We note that while the hotter gas is dominating this region due to it occupying a large number of voxels, the density of the collisionally ionized hotter gas is lower, and this region may be less observationally relevant than the regions which follow the \citet{madsen06} assumptions. Other assumptions made by \citet{madsen06} which differ from these full 3D time-dependent models include; the ionized fraction in the WIM is near unity, little nitrogen is ionized to \ion{N}{III} (due to no collisional ionization), a constant $N^{+}/N$ ratio of 0.8.
The $S^{+}/S$ contours also follow the expected trends, but the simulation data shows the transition in ionization state happening slightly faster across the diagnostic diagram than shown in \citet{madsen06}. However, discrepancies between the observational theory and simulated results are slight, and in general the use of these diagnostic diagrams as a measure of temperature and ionization state of the gas appears to be robust.

\section{Discussion}

\subsection{Comparisons to Benchmarks}

We have developed \texttt{CMacIonize} to further explore and understand the origins of various emission line signatures of the DIG. By benchmarking our new code against \texttt{CLOUDY} in section~\ref{benchmarking}, we have ensured that the time-dependent ionization calculations produce the correct ionization and temperature structure at the end of the development of the ionized region. The good agreement between the \texttt{CLOUDY} and \texttt{CMacIonize} line intensities and other metrics shows that our code is appropriate for the DIG temperature regime.

Running the full radiation-hydrodynamics code with tabulated cooling rates for the Lexington benchmarks shows large discrepancies in ionization state, temperature, size of ionized region and ultimately line intensities. The lack of dependence of the cooling rates on the ionization state of the gas gives cooling rates which are too low for the ionized region, resulting in higher temperatures and a larger ionized region. The temperature does not display the same radial structure present in the simulation with the the full cooling code, suggesting that the temperature structure of the ionized region is controlled not only by radiation hardening (whereby photons of higher energies travel further than lower energy photons), but the ionization structure of each cooling ion. This is an effect that will occur not only on the sizescale of an H{\sc II} region, but also on kpc scales to explain rising emission line ratios (and temperatures) with height above the midplane.

\subsection{Full 3D ionization structure}

In \citet{mccallum24}, the time-dependent ionization calculation generates a DIG layer which is more extended and less time variable than simulations adopting ionization equilibrium. Our new simulations show that this result is also common to other ionized species such as \ion{N}{II}  and \ion{O}{II} . With a greater extent of ionized hydrogen, the optical depth to Lyman continuum photons is lower, and ionizing photons are able to travel further and higher in the galactic disc. This produces a more extended layer of \ion{N}{II} , \ion{O}{II}  and other ions of similar ionization energy to hydrogen.

The effect of the time-dependence increasing the spatial extent of ionized species is especially apparent in photoionized species with high ionization energies such as \ion{Ne}{III} . The equilibrium assumption only allows for rare high energy photons to ionize the regions immediately surrounding the ionizing sources. With lower optical depths to high altitude in the time-dependent simulation, we find \ion{Ne}{III}  being formed at heights above 2~kpc. This highly ionized neon is able to persist for longer in regions of lower density, giving a height dependence of the fraction of \ion{Ne}{III} . This density (height) dependence gives strongly rising [NeIII]/[NeII] emission line ratios which has previously been interpreted as a hardening of the radiation field with height due to a transition from OB ionized gas to HOLMES ionized gas (see \citet{rand08}, \citet{floresfjardo}). Our simulations suggest that these rising ratios of [NeIII]/[NeII] can be explained without a non-OB secondary (HOLMES) source of ionization simply due to the disc geometry and time-dependent effects.

\subsection{Line ratios versus observations}

Our \emph{fiducial} simulation produced emission line ratios similar to structures observed in NGC~891. The largest discrepancies are the elevated values of [SII]/H$\alpha$ and [NII]/H$\alpha$ at heights of less than 1~kpc. However the general trend of increasing [SII]/H$\alpha$ and [NII]/H$\alpha$ with height above the midplane is seen. The heightened [NII]/H$\alpha$ and [SII]/H$\alpha$ line ratios in the midplane could be due to differences in analysing a full edge on galaxy compared to the restricted $\rm{kpc}^{2}$ patch of galaxy of our simulations. The elevated line ratios could also suggest the WIM at heights of $|z| < 1 \rm{kpc}$ is hotter than reality. This could be due to a variety of factors, such as too high a supernova rate or ionizing luminosity. The simulations of \citet{mccallum24} were set up to replicate the Milky Way's vertical structure, but extra missing dynamical and ionization support mechanisms such as magnetic fields or cosmic ray ionization could lead to an overestimation of the feedback mechanisms.

The work of \citet{rand08}, which provided the NGC~891 data included in figure~\ref{lineratios}, concludes that the high values of [NeIII]/[NeII] which increase with height could be evidence for a secondary population of hot ionising sources. This conclusion stems from the rarity of photons energetic enough to ionize \ion{Ne}{II}  to \ion{Ne}{III}  in even the hottest OB star spectra. We find that our \emph{equilibrium} simulation supports the idea that midplane OB stars are unable to produce an extended layer of \ion{Ne}{III} , with the small amount of [Ne III] 15~$\mu \rm{m}$ emission confined only to the midplane regions very close to the ionizing sources. We find however that the inclusion of a full time-dependent ionization scheme alleviates the need for a large abundance of hard photons to explain the trends in [Ne III]/[Ne II] with height, and that the observational results from NCG~891 are replicated in our \emph{fiducial} simulation without a secondary population of HOLMES sources. The long recombination times in the high altitude gas means that the occasional exposure to the rare photons with energies above 41.0~eV is sufficient to keep the layer of \ion{Ne}{III}  in its high state of overionization compared to ionization equilibrium. Other line ratios are replicated in value and trend with height, with the exception of [OI]/H$\alpha$, which is approximately a factor of 2 higher than observed.

The full radiation-hydrodynamics \emph{table cooling} simulation, adopting time-dependent ionization but tabulated cooling rates produces line ratio structures with height which are broadly in agreement with the high altitude gas in the \emph{fiducial} simulation. This suggests that the tabulated cooling curves are sufficient for lower density gas in which photoionization is not necessarily the dominant source of ionization. However, we see an excess of [NII]/H$\alpha$ and [SII]/H$\alpha$ in the midplane, which dominates the full grid integrated line ratios due to the high emission measure in these regions. The \emph{table cooling} run produces a global [NII]/H$\alpha$ of 1.27 versus the fiducial value of 0.54, caused by an excess of hotter, dense, photoionized midplane gas. This agrees with the table cooling benchmark tests, whereby the tabulated cooling rates were too low to represent realistic cooling in ionized gas. This suggests that midplane DIG temperatures in the work of \citet{mccallum24} are also likely too high. The conclusions of this previous work however are robust to a 50\% increase in DIG temperature. This extent of change in DIG temperature however has large impacts on synthetic emission line intensities, further motivating the need for the full cooling calculation included the other three simulations (\emph{fiducial}, \emph{equilibrium} and \emph{HOLMES}).

While the \emph{table cooling} simulation is unreliable for the generation of synthetic emission maps from the denser midplane DIG, the \emph{equilibrium} run successfully reproduces the \emph{fiducial} results for the midplane. However, the line ratios become more discrepant at higher altitudes and lower densities. This is due to the time-dependent effects being density dependant. \citet{mccallum24} concluded that ionization equilibrium was a good assumption when the timescale for the variability of the ionization is longer than the recombination timescale of the gas. This criteria holds true for the denser midplane gas with shorter recombination times, but not for the low density, high altitude DIG. Figure~\ref{ratiovhalpha} demonstrates that the equilibrium run departs from the fiducial line ratio results in an H$\alpha$ surface brightness of approximately 1~R. The difficulty in reproducing reliable line ratios in low density gas with the assumption of ionization equilibrium is also seen in figure~\ref{lineratios}, whereby high altitude line ratios from NCG~891 are matched well in the time-dependent runs, but not with equilibrium ionization.

\subsection{The role of HOLMES}

\begin{figure}
    \centering
    \includegraphics[width=\columnwidth]{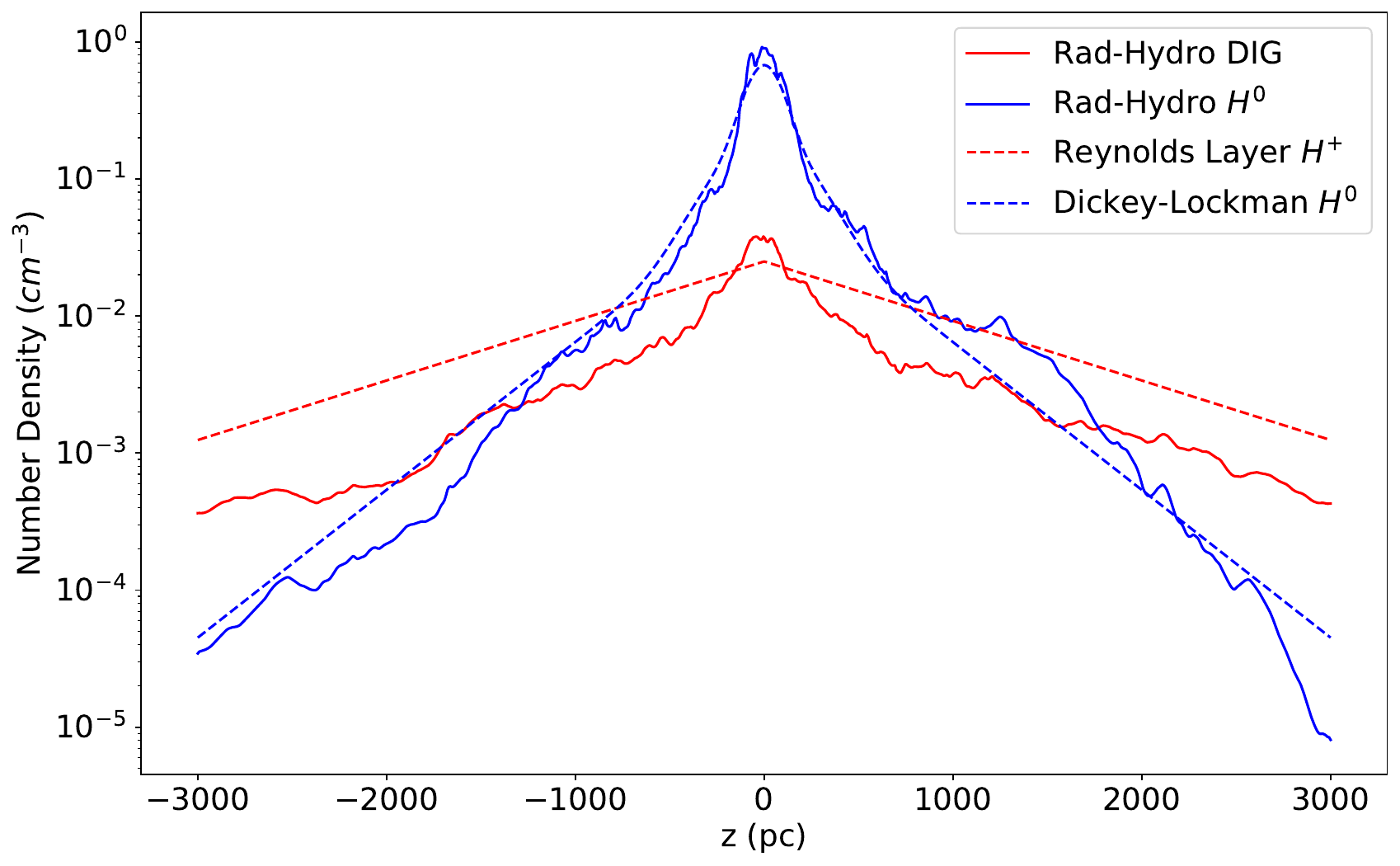}
    \caption{Vertical extent of both neutral and ionized hydrogen in the \emph{fiducial} simulation after 25~Myr of evolution. Also shown are the Dickey-Lockman approximation for neutral hydrogen distribution and Reynolds layer estimate for vertically extended DIG. As seen in \citet{mccallum24}, there is a general deficit of DIG in the simulations without HOLMES sources, although the scale height is comparable to observational estimates.}
    \label{vertstruct}
\end{figure}

The closest match to observed vertically resolved line ratios comes from our \emph{fiducial} simulation. It is noted in \citet{mccallum24} that the closest match to the total ionized hydrogen density structure requires a population of 1000 HOLMES sources, each with a luminosity of $5\times10^{45} {\rm s}^{-1}$. This population of HOLMES sources is included in our \emph{HOLMES} run, and further ionizes hydrogen due to their vertical extent. However, the influence of such a population on the emission line ratios moves the results further from observed values due to the hard spectra of the HOLMES sources. The luminosity and spatial density of these sources is not well determined. In \citet{mccallum24} the luminosity of these sources was treated as a free parameter in order to match the Reynolds layer of ionized hydrogen. Our new metal ionization results suggest that the previous best fit luminosity may be too high to allow for reasonable line ratios amongst such a hard radiation field. Further modifications to the hydrogen only simulations such as varying the ionizing photon molecular cloud escape fraction or including cosmic ray ionization could replace HOLMES sources as a source of additional ionization to reproduce the Reynolds layer of the DIG. The extent of the DIG layer produces in our fiducial model is shown in figure~\ref{vertstruct}, where the density of the DIG is seen to be lower than observed, but with a comparable scaleheight.

\subsection{Comparison to other simulations}

The simulations of \citet{vandenbroucke18} and \citet{vandenbroucke19} utilise the same hydro scheme as this work, and similarly the cooling rates and line emissivities are calculated in the same way. The key modifications between these publications and this one is the development of the non-equilibrium temperature and ionization state calculator, the effects of which are documented in the work of \citet{mccallum24} and this paper. We have also updated the atomic data to use the CHIANTI database from \citet{delzanna21}. This has a modest effect on the structures of most ions, with the exception of sulphur which showed larger changes from the our previous dataset. The collisional excitation rates for the emission lines remains the same as was benchmarked in \citet{cmi2}.

The most advanced tall box simulations for the study of the dynamics of the ISM remains those of the SILCC project (\citet{walch15}, \citet{rathjen21}, \citet{rathjen23}) and TIGRESS-NCR (\citet{kim23}). These simulations contain physics which is not included in our setup, most notably the inclusion of gas self-gravity for the self-consistent simulation of star formation through sink particles, magnetic fields which have been shown to provide vertical support to the ISM, and cosmic rays which may be a further contributor to the ionization of the high altitude gas. Acknowledging these missing aspects in our simulations, we have focused our efforts on the reliable calculation of DIG temperature and ionization state. Our simulations are among the first to introduce the full time-dependent calculation of the metal ionization states. 

We have shown that simplifying assumptions such as ionization equilibrium and the use of a 1D cooling curve can be effective under certain physical conditions of the ISM. \citet{mcclymont24} utilised a post-processing technique on an isolated galaxy simulation to calculate various emission lines. Convincing matches of simulation results to observational data from MUSE are found for line ratios as a function of H$\alpha$ surface brightness (as in our figure~\ref{ratiovhalpha}), without the need for a full calculation of each ionization state as a function of its ionization state history. We note that we do not have cells with $H\alpha$ surface brightness as high as those of \citet{mcclymont24}, likely due to our lower midplane resolution. With \citet{mcclymont24} focusing on the face-on emission from better resolved, and hence denser midplane gas, they are simulating a regime where the assumption of ionization equilibrium is robust (see the convergence of the \emph{equilibrium} and \emph{fiducial} simulations in figure~\ref{ratiovhalpha} towards higher values of H$\alpha$ surface brightness.)

Another code with the full time-dependent calculation of such a number of coolants is described in \citet{katz22}, who have developed the RAMSES-RT code to include these processes. While these simulations have been developed with focus towards cosmological scales, we believe the RAMSES-RT code could also be used to simulate the vertically extended low density DIG.

\section{Conclusions}

We have extended the simulations of \citet{vandenbroucke19} and \citet{mccallum24} to more reliably model the galactic DIG not just in hydrogen dynamics and ionization, but also in the ionization of a number of metal coolants. We have included the full time-dependent ionization and recombination of a number of ionized species. As was found in \citet{mccallum24} for hydrogen ionization, high altitude DIG metal ionization also exhibits a dependence upon its ionization history. This suggests that the assumption of ionization equilibrium is not appropriate for the accurate modelling of high altitude DIG emission lines.

The cooling algorithm has been improved to calculate cooling rates self-consistently in the DIG as a function of the ionization state of each coolant. This improves our previous simulations which used tabulated cooling rates as a function of temperature only.

The new code has been benchmarked against the equilibrium ionization code \texttt{CLOUDY}, and the time-dependent evolution of each benchmark was found to evolve to the \texttt{CLOUDY} solution for the ionization structure of each ion, and the temperature structure. Turning off the full cooling calculation and using the original tabulated cooling curves gives inaccurate temperature structures and emission line intensities.

The \emph{fiducial} 3D tall box simulation produce a set of vertically resolved line ratios which agree with published data from NGC~891. The NRMSE is calculated for low and high altitude gas for each line ratio. Good agreement (NRMSE < 0.5) is found globally for [NeIII]/[NeII], [HeI]/$H\alpha$ and [OIII]/$H\beta$, and in the high altitude [NII]/$H\alpha$ and [SII]/$H\alpha$. The global ratios of [OI]/H$\alpha$ are overpredicted to a large extent (NRMSE > 7). The \emph{equilibrium} run fails to produce comparable ionization state structures, and the \emph{table cooling} run fails to produce a realistic DIG temperature structure in the photoionized gas of densities above 0.1 $\rm{cm}^{-3}$, resulting in unrealistic line intensities. 

The results of our tall box simulations suggest that in order to accurately model emission lines from the high altitude DIG, the accurate non-equilibrium ionization state of a large number of coolants must be known. The simplifying assumption of ionization equilibrium works well for higher density gas, but not for high altitude, low density DIG. We find that summing the total H$\alpha$ in our \emph{fiducial} and \emph{equilibrium} runs above a density cut off of $5.9~\rm{cm}^{-3}$ produces the same total emissivity. This required cut-off will however be dependent on variables such as the stochasticity and time-variability of the ionizing sources. The application of a time saving 1D tabulated cooling curve produces reliable DIG temperatures (and hence line ratios) in lower density cells, but produces DIG which is too hot near the midplane where the densities are higher and photoionization dominates the ionization.

The high altitude time-dependent build-up of high ionization energy ions also calls for caution when interpreting emission lines from ions with high ionization potentials. Our \emph{fiducial} simulation produces elevated [NeIII]/[NeII] ratios with only an OB-star spectrum. Elevated values of this ratio has previously been interpreted as evidence of a secondary source of ionization with a hard spectrum, but can be recreated with a softer 40kK OB spectra combined with time-dependent effects.

The tall box simulation, \emph{HOLMES}, containing a population of hot low mass evolved stars produces line ratios that do not match those observed in the Milky Way or NGC~891. We interpret this as evidence that the HOLMES luminosity we used is higher than the true value, or that the HOLMES spectrum we used is too "hard". We have recreated a realistic set of high altitude line ratios using only OB stars as the source of ionization. While there is still a small discrepancy in the ionization structure compared to observations without the HOLMES population, our simulations suggest that it difficult to obtain this extra ionized gas by HOLMES ionization, without significantly shifting the computed line ratios from observed values.

\section*{Acknowledgements}

LM acknowledges financial support from a UK-STFC PhD studentship. DK acknowledges support from an HST Archival Theory Program 17060, provided by NASA through a grant from the Space Telescope Science Institute, which is operated by the Association of Universities for Research in Astronomy, Inc., under NASA contract NAS5-26555.
\section*{Data Availability}

Any underlying data from this work will be made available upon reasonable request.



\bibliographystyle{mnras}
\bibliography{biblio} 




\appendix

\section{Benchmark Line Intensities}

The total intensity of each emission line for the various Lexington benchmarking runs were calculated in $\rm{erg}~\rm{s}^{-1}$ and normalised to the intensity of the $H\beta$ line.

The \texttt{CMacIonize} values (CMI) are from the full time-dependent ionization code after the ionization front has stopped evolving and is hence in ionization equilibrium. CMItable shows the equivalent results using the \citet{derijke13} cooling curves. \texttt{CLOUDY} equilibrium results are also shown.

Values for the HII20 and HII40 benchmarks are displayed. Lines are the same as shown in \citet{cmacionize}. Other values displayed are the total $H\beta$ luminosity, the temperature at the surface of the evacuated region in the density setup, the $n_{e}n_{H^{+}}$ weighted temperature and the outer radius of the ionized region.

\begin{table}
\centering
\resizebox{\columnwidth}{!}{%
\begin{tabular}{lccc}
\hline
\textbf{HII20}                                & \textbf{CLOUDY} & \textbf{CMI} & \textbf{CMItable} \\ \hline
H$\beta$ luminosity $(\rm{erg}~\rm{s}^{-1})$ & $4.99\times10^{36} $        & $4.86\times10^{36}$     & $4.61\times10^{36}$          \\
$T_{inner} ~(\rm{K})$                                   & 7155            & 6954         & 10419             \\
$\langle T[n_{e}n_{H^{+}}]\rangle ~(\rm{K})$            & 6910            & 6962         & 10213             \\
$R_{outer} ~(\rm{m})$                                   & $8.90\times10^{16}$        & $8.85\times10^{16}$    & $9.77\times10^{16}$        \\ \cline{1-4}
{[CII] 2325~\AA\ multiplet}                   & 0.0756          & 0.0814      & 1.29           \\
{[NII] 122~$\mu$m}                            & 0.0696          & 0.0696      & 0.0939           \\
{[NII] 6584~\AA\ and 6548~\AA}                & 0.809          & 0.902   & 2.90           \\
{[NII] 5755~\AA}                              & 0.0032          & 0.00329      & 0.0307            \\
{[NIII] 57.3~$\mu$m}                          & 0.0038          & 0.00354      & 0.00505           \\
{[OI] 6300~\AA\ and 6363~\AA}                 & 0.0083          & 0.0178      & 0.0172           \\
{[OII] 7320~\AA\ and 7330~\AA}                & 0.0113          & 0.0120      & 0.178           \\
{[OII] 3726~\AA\ and 3729~\AA}                & 1.40          & 1.48      & 9.91           \\
{[OIII] 51.8~$\mu$m}                          & 0.0015          & 0.00153      & 0.00231           \\
{[OIII] 88.3~$\mu$m}                          & 0.0018          & 0.0018       & 0.00278           \\
{[OIII] 5007~\AA\ and 4959~\AA}               & 0.0017          & 0.00214      & 0.0130           \\
{[NeII] 12.8~$\mu$m}                          & 0.296          & 0.298       & 0.372           \\
{[SII] 6716~\AA\ and 6731~\AA}                & 0.744          & 0.792      & 2.21           \\
{[SII] 4068~\AA\ and 4076~\AA}                & 0.0226          & 0.0237      & 0.117           \\
{[SIII] 18.7~$\mu$m}                          & 0.209           & 0.212      & 0.303            \\
{[SIII] 33.6~$\mu$m}                          & 0.341          & 0.355      & 0.471           \\
{[SIII] 9532~\AA\ and 9069~\AA}               & 0.293          & 0.312      & 0.863           \\ \hline
\end{tabular}%
}
\caption{Resulting line intensities and other characteristics from the HII20 Lexington benchmark in \texttt{CLOUDY}, \texttt{CMacIonize} in time dependent mode, and \texttt{CMacIonize} in time dependent mode but using cooling rates from the tabulated cooling curves of \citet{derijke13}. Line intensities are normalised total emissivity to the total emissivity of H$\beta$.}
\label{tab:my-table}
\end{table}

\begin{table}
\centering
\resizebox{\columnwidth}{!}{%
\begin{tabular}{lccc}
\hline
\textbf{HII40}                                & \textbf{CLOUDY} & \textbf{CMI} & \textbf{CMItable} \\ \hline
H$\beta$ luminosity $(\rm{erg}~\rm{s}^{-1})$ & $2.06\times10^{37}$ & $2.03\times10^{37}$ & $1.98\times10^{37}$ \\
$T_{inner}~(\rm{K})$                         & 7839              & 7535           & 11179           \\
$\langle T[n_{e}n_{H^{+}}]\rangle~(\rm{K})$  & 8240              & 8155           & 11096           \\
$R_{outer}~(\rm{m})$                         & $1.46\times10^{17}$ & $1.44\times10^{17}$ & $1.57\times10^{17}$ \\ \cline{1-4}
{[HeI] 5876~\AA}                              & 0.101             & 0.118           & 0.106            \\
{[CII] 2325~\AA multiplet}                              & 0.191             & 0.189           & 0.692            \\
{[CIII] 1907~\AA\ and 1909~\AA}                 & 0.064             & 0.0806          & 1.34             \\
{[NII] 122~$\mu$m}                            & 0.0296            & 0.0280          & 0.0359           \\
{[NII] 6584~\AA\ and 6548~\AA}                  & 0.672             & 0.688           & 1.29             \\
{[NII] 5755~\AA}                              & 0.0061            & 0.00555         & 0.0166           \\
{[NIII] 57.3~$\mu$m}                          & 0.285             & 0.317           & 0.382            \\
{[OI] 6300~\AA\ and 6363~\AA}                   & 0.0112            & 0.0249          & 0.0163           \\
{[OII] 7320~\AA\ and 7330~\AA}                  & 0.0336            & 0.0311          & 0.130            \\
{[OII] 3726~\AA\ and 3729~\AA}                  & 2.34              & 2.18            & 6.23             \\
{[OIII] 51.8~$\mu$m}                          & 1.22              & 1.25            & 1.49             \\
{[OIII] 88.3~$\mu$m}                          & 1.08              & 1.48            & 1.79             \\
{[OIII] 5007~\AA\ and 4959~\AA}                 & 2.52              & 2.60            & 9.94             \\
{[OIII] 4363~\AA}                             & 0.0048            & 0.00468         & 0.0649           \\
{[NeII] 12.8~$\mu$m}                          & 0.200             & 0.181           & 0.213            \\
{[NeIII] 15.5~$\mu$m}                         & 0.217             & 0.327           & 0.393            \\
{[NeIII] 3869~\AA\ and 3968~\AA}                & 0.0595            & 0.0906          & 0.458            \\
{[SII] 6716~\AA\ and 6731~\AA}                  & 0.306             & 0.350           & 0.584            \\
{[SII] 4068~\AA\ and 4076~\AA}                  & 0.0143            & 0.0159          & 0.0340           \\
{[SIII] 18.7~$\mu$m}                          & 0.506             & 0.496           & 0.642            \\
{[SIII] 33.6~$\mu$m}                          & 0.792             & 0.801           & 0.974            \\
{[SIII] 9532~\AA\ and 9069~\AA}                 & 1.01              & 0.992           & 2.02             \\
{[SIV] 10.5~$\mu$m}                           & 0.190             & 0.197           & 0.207            \\ \hline
\end{tabular}%
}
\caption{Resulting line intensities and other characteristics from the HII40 Lexington benchmark in \texttt{CLOUDY}, \texttt{CMacIonize} in time dependent mode, and \texttt{CMacIonize} in time dependent mode but using cooling rates from the tabulated cooling curves of \citet{derijke13}. Line intensities are normalised total emissivity to the total emissivity of H$\beta$.}
\label{tab:my-table-2}
\end{table}


\bsp	

\label{lastpage}
\end{document}